\documentclass[journal]{vgtc}                

\ifpdf
  \pdfoutput=1\relax                   
  \usepackage{graphicx}                
  \DeclareGraphicsExtensions{.pdf,.png,.jpg,.jpeg} 
\else
  \usepackage{graphicx}                
  \DeclareGraphicsExtensions{.eps}     
\fi%

\graphicspath{{figures/}{pictures/}{images/}{./}} 

\usepackage{microtype}                 
\PassOptionsToPackage{warn}{textcomp}  
\usepackage{textcomp}                  
\usepackage{mathptmx}                  
\usepackage{times}                     
\usepackage{cite}                      
\usepackage{tabu}                      
\usepackage{booktabs}                  

\usepackage{gensymb}
\usepackage{threeparttable}
\usepackage{multirow}
\usepackage{warpcol}
\usepackage{makecell}
\usepackage{amsthm,amsmath,amssymb}
\usepackage{mathrsfs}
\newcommand{\etal}{\textit{et al}.}
\newcommand{\ie}{\textit{i}.\textit{e}.}
\newcommand{\eg}{\textit{e}.\textit{g}.}
\newcommand{\tabincell}[2]{\begin{tabular}{@{}#1@{}}#2\end{tabular}}
\usepackage{wrapfig}
\usepackage{subfig}
\usepackage{color,xcolor}

\vgtccategory{Research}
\vgtcpapertype{please specify}

\title{Perceptual Quality Assessment of Omnidirectional \\Images as Moving Camera Videos}
\author{Xiangjie Sui, Kede Ma, \textit{Member, IEEE}, Yiru Yao, and Yuming Fang, \textit{Senior Member, IEEE}}
\authorfooter{
\item
Xiangjie Sui, Yiru Yao, and Yuming Fang are with the School of Information Management, Jiangxi University of Finance and Economics, Nanchang 330032, Jiangxi, China (e-mail: suixiangjie2017@163.com, 2848444870@qq.com, fa0001ng@e.ntu.edu.sg).
\item
Kede Ma is with the Department of Computer Science, City University of Hong Kong, Kowloon, Hong Kong (e-mail: kede.ma@cityu.edu.hk).
\
}

\shortauthortitle{Biv \MakeLowercase{\textit{et al.}}: Global Illumination for Fun and Profit}

\abstract{Omnidirectional images (also referred to as static 360{\degree} panoramas) impose viewing conditions much different from those of regular 2D images. How do humans perceive image distortions in immersive virtual reality (VR) environments is an important problem which receives less attention. We argue that, apart from the distorted panorama itself, two types of VR viewing conditions are crucial in determining the viewing behaviors of users and the perceived quality of the panorama: the starting point and the exploration time. We first carry out a psychophysical experiment to investigate the interplay among the VR viewing conditions, the user viewing behaviors, and the perceived quality of 360{\degree} images. Then, we provide a thorough analysis of the collected human data, leading to several interesting findings. Moreover, we propose a computational framework for objective quality assessment of 360{\degree} images, embodying viewing conditions and behaviors in a delightful way. Specifically, we first transform an omnidirectional image to several video representations using different user viewing behaviors under different viewing conditions. We then leverage advanced 2D full-reference video quality models to compute the perceived quality. We construct a set of specific quality measures within the proposed framework, and demonstrate their promises on three VR quality databases.}


\teaser{
  \centering
  \includegraphics[width=\linewidth]{Graphical_Abstract.pdf}
  \caption{Illustration of how humans explore static omnidirectional panoramas. Each user may have different viewing behaviors (\ie, scanpaths) under different viewing conditions, giving rise to different video representations of the same 360{\degree} image with varying perceived quality. We consider two types of viewing conditions - the starting point and the exploration time. Specifically, the starting point provides the longitude and the latitude, where the initial viewport can be extracted. A gaze scanpath is generated, when each user is freely exploring the virtual environment within the exploration time. A video sequence that contains only global motion can then be obtained by sampling, along each user's scanpath, a number of viewports from the omnidirectional panorama. We compute the perceived quality of the omnidirectional image by comparing it to its reference using existing video quality measures.}
	\label{fig:teaser}
}

\vgtcinsertpkg

\begin{document}


\firstsection{Introduction}

\maketitle
Virtual reality (VR) photography is the art of capturing or creating a complete natural scene as a single omnidirectional  image~\cite{Wikipedia_Intro}, also known as a static 360{\degree} panorama. The viewing experience enabled by omnidirectional images is substantially different from traditional multimedia data, as humans are allowed to freely explore immersive virtual environments (see \autoref{fig:teaser}). Therefore, understanding how humans perceive visual distortions of omnidirectional images emerges as a new research direction due to its importance to panoramic image acquisition, compression, storage, transmission, and reproduction~\cite{XU2020Review_Intro}.

\begin{figure}[t]
 \centering
 \includegraphics[width=\columnwidth]{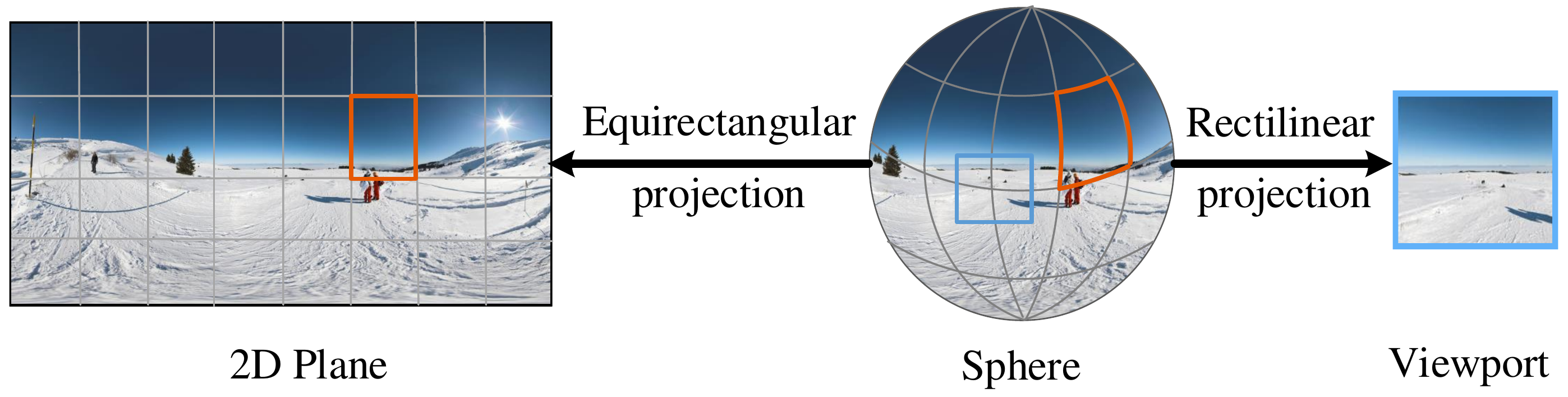}
 \caption{Illustration of different representations of an omnidirectional image. Equirectangular projection is commonly used to obtain a 2D plane for storage, while rectilinear projection is adopted to extract a viewport for visual consumption at a time instant.}
 \label{fig:sphere2plane}
\end{figure}

Objective quality assessment of omnidirectional images is often performed in 2D projected planes by leveraging existing 2D image quality assessment (IQA) models (see Fig.~\ref{fig:sphere2plane}). However, different map projections come with different problems. For example, equirectangular projection generates severe shape distortions near the poles, whereas cube map projection has an oversampling rate of up to $190\%$ compared to the sphere~\cite{King2005Compression_Intro}. It follows that distortions measured in the 2D plane may not correspond to distortions observed in the sphere. To alleviate the projection mismatch, several objective IQA models~\cite{Sun2017WSPSNR_OIQA,Lopes2018WSSIM_OIQA} make local quality measurements in the plane, and pool them using spherical areas as weightings. A better implementation of this idea is to compute quality estimates uniformly over the sphere~\cite{Yu2015SPSNR_OIQA,Chen2018SSIM_OIQA}.

 In 2D IQA, user viewing behavior can be well controlled in a laboratory environment, and is often assumed similar without explicit modeling. However, this assumption does not hold in omnidirectional IQA. Equipped with a head-mounted display (HMD), humans are able to use both significant head and gaze movements to explore viewports of interest in the scene. Recently, Sitzmann~\etal~\cite{Sitzmann2018Saliency_Analyze} found that under different viewing conditions, agreement among gaze scanpaths of subjects is not high. To the best of our knowledge, no existing work gives a complete treatment of viewing behavior when evaluating the perceived quality of omnidirectional images.

\begin{table*}[!t]
    \small
    \renewcommand\tabcolsep{4pt}
    \centering
    \begin{tabular}{c|c c c c c c c}
    \toprule
         Database & Projection & \# of images & Resolution & Exploration time & HM/EM data  & Distortion type &  Public availability \\ \hline

         Upenik \etal \cite{Upenik2016_subjective} & \tabincell{c}{ERP\\CMP} & 6/54 & \tabincell{c}{3,000$\times$1,500 (ERP)\\ 2,250$\times$1,500 (CMP)} & 30 & HM & \tabincell{c}{JPEG compression, Projection} & N/A  \\ \hline

         Duan \etal \cite{Duan2018_subjective} & ERP & 16/320 & \tabincell{c}{ 11,332$\times$5,666\\to 13,320$\times$6,660} & 20 & HM and EM & \tabincell{c}{JPEG compression, JPEG2000 compression,\\ Gaussian blur, Gaussian noise} & Upon request \\ \hline

        Sun \etal \cite{Sun2017_subjective} & ERP & 16/528 & 4,096$\times$2,048 & N/A & N/A & \tabincell{c}{JPEG compression,\\H.264 compression, H.265 compression}  & N/A \\ \hline

        Huang \etal \cite{Huang2018_subjective} & ERP & 12/144 & 4,096$\times$2,160 & 20 & N/A & \tabincell{c}{Downsampling, JPEG compression} & \tabincell{c}{https://vision.nju.edu.\\cn/20/87/c29466a467\\079/page.htm} \\ \hline

        Chen \etal \cite{Chen2020_subjective} & ERP & 15/450 & 4,096$\times$2,048 & 20 & HM and EM & \tabincell{c}{Gaussian noise, Gaussian blur,\\Downsampling, Stitching distortion,\\VP9 compression, H.265 compression} & \tabincell{c}{http://live.ece.utexas.\\edu/research/VR3D/\\index.html} \\ \hline

        Ours & ERP & 36/72 & 7,680$\times$3,840 & 5/15 & HM and EM & Stitching distortion, H.265 compression & \tabincell{c}{https://github.com/\\xiangjieSui/img2video}\\
    \bottomrule
    \end{tabular}
    \caption{Summary of VR IQA databases. ERP and CMP stand for the equirectangular projection and the cube map projection, respectively. The data in the ``\# of images'' column is in the  form of ``\# of reference images/\# of distorted images.''}
    \label{tab:databaes}
\end{table*}

In this paper, we argue that there are at least two types of VR viewing conditions -- the starting point and the exploration time -- that significantly affect viewing behavior, and subsequently determine the perceived VR quality. The viewing behavior, represented by the so-called scanpath, is a 2D gaze trajectory over the sphere~\cite{Noton1971Scanpath_Intro}. The starting point provides the longitude and the latitude, at which the initial viewport is centered (see Fig.~\ref{fig:sphere2plane}). The exploration time records how long it takes for a user to finish exploring an omnidirectional image.

We first conduct a psychophysical experiment to study the interplay among the VR viewing conditions, the user viewing behaviors, and the perceived quality of panoramic images.  Thorough analysis of the collected human data validates that viewing conditions have an important impact on the perceived quality of omnidirectional images.
Furthermore, we propose a computational framework for objective quality assessment of distorted panoramas, incorporating viewing conditions and behaviors. Specifically, we represent a panorama by moving camera videos, where we sample, along different users' scanpaths, sequences of rectilinear projections of viewports. The resulting videos contain only global motion~\cite{Dufaux2000Globalmotion_Intro}, as if they were captured by a moving camera where the moving patterns are determined by user viewing behaviors. Instead of learning omnidirectional IQA models from scratch, the novel video representations allow us to directly adapt existing video quality assessment (VQA) tools to this immersive application. We construct several quality models within the proposed computational framework by first predicting frame-level quality using existing 2D IQA models~\cite{Wang2006Book_Intro} and then pooling the quality estimates temporally~\cite{Tu2020Compare_VQA}. Extensive experiments on the proposed database and two publicly available VR databases~\cite{Duan2018_subjective,Chen2020_subjective} demonstrate the promise of our framework for objective quality assessment of 360{\degree} images.

\begin{figure*}[!t]
    \centering
    \captionsetup{justification=centering}
    \subfloat{\includegraphics[width = 0.155\linewidth]{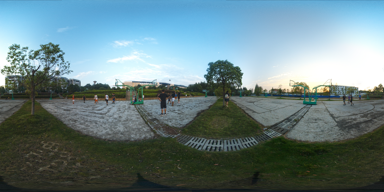}}\hskip.2em
    \subfloat{\includegraphics[width = 0.155\linewidth]{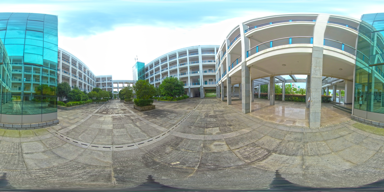}}\hskip.2em
    \subfloat{\includegraphics[width = 0.155\linewidth]{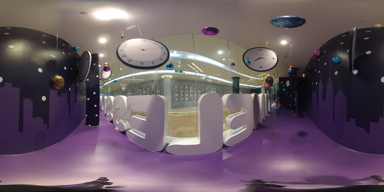}}\hskip.2em
    \subfloat{\includegraphics[width = 0.155\linewidth]{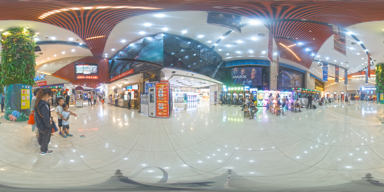}}\hskip.2em
    \subfloat{\includegraphics[width = 0.155\linewidth]{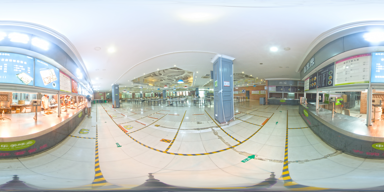}}\hskip.2em
    \subfloat{\includegraphics[width = 0.155\linewidth]{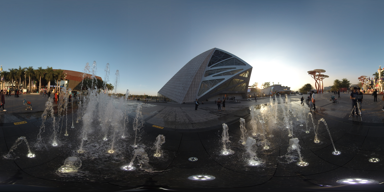}}\hskip.2em

    \subfloat{\includegraphics[width = 0.155\linewidth]{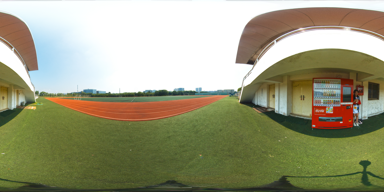}}\hskip.2em
    \subfloat{\includegraphics[width = 0.155\linewidth]{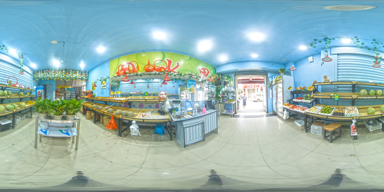}}\hskip.2em
    \subfloat{\includegraphics[width = 0.155\linewidth]{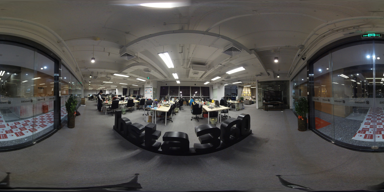}}\hskip.2em
    \subfloat{\includegraphics[width = 0.155\linewidth]{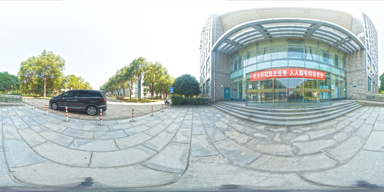}}\hskip.2em
    \subfloat{\includegraphics[width = 0.155\linewidth]{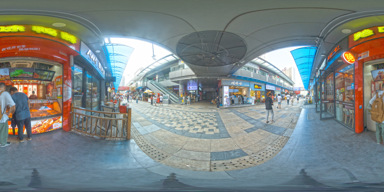}}\hskip.2em
    \subfloat{\includegraphics[width = 0.155\linewidth]{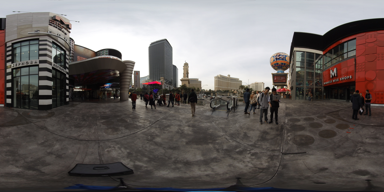}}\hskip.2em

    \subfloat{\includegraphics[width = 0.155\linewidth]{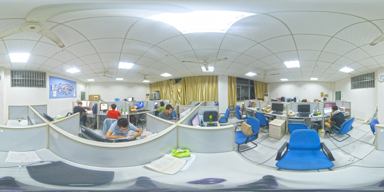}}\hskip.2em
    \subfloat{\includegraphics[width = 0.155\linewidth]{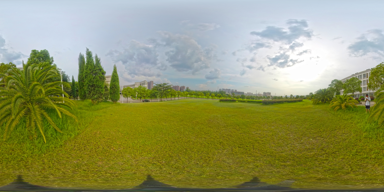}}\hskip.2em
    \subfloat{\includegraphics[width = 0.155\linewidth]{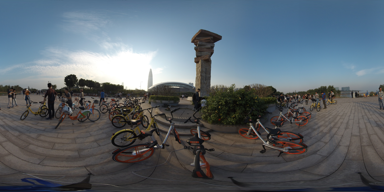}}\hskip.2em
    \subfloat{\includegraphics[width = 0.155\linewidth]{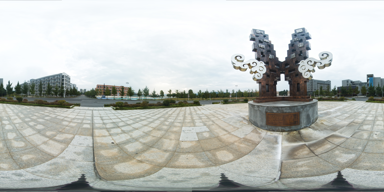}}\hskip.2em
    \subfloat{\includegraphics[width = 0.155\linewidth]{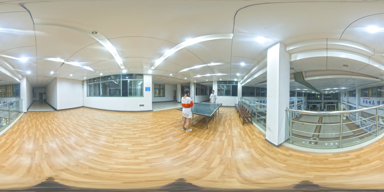}}\hskip.2em
    \subfloat{\includegraphics[width = 0.155\linewidth]{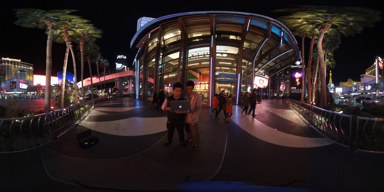}}\hskip.2em

    \subfloat{\includegraphics[width = 0.155\linewidth]{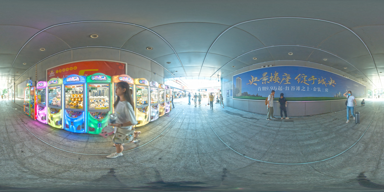}}\hskip.2em
    \subfloat{\includegraphics[width = 0.155\linewidth]{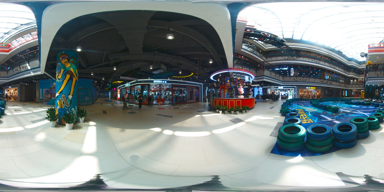}}\hskip.2em
    \subfloat{\includegraphics[width = 0.155\linewidth]{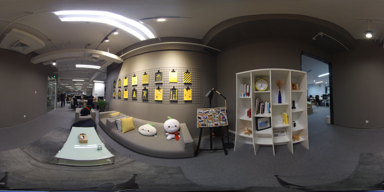}}\hskip.2em
    \subfloat{\includegraphics[width = 0.155\linewidth]{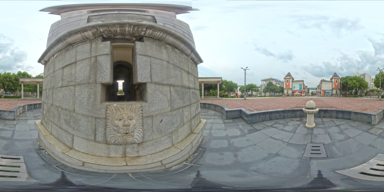}}\hskip.2em
    \subfloat{\includegraphics[width = 0.155\linewidth]{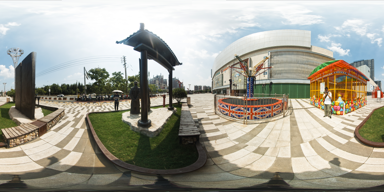}}\hskip.2em
    \subfloat{\includegraphics[width = 0.155\linewidth]{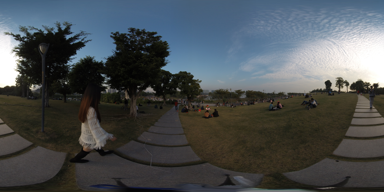}}\hskip.2em

    \subfloat{\includegraphics[width = 0.155\linewidth]{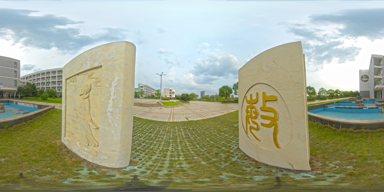}}\hskip.2em
    \subfloat{\includegraphics[width = 0.155\linewidth]{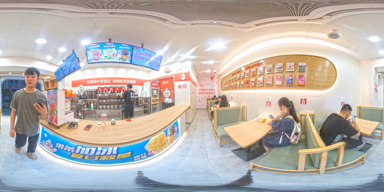}}\hskip.2em
    \subfloat{\includegraphics[width = 0.155\linewidth]{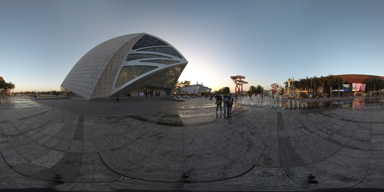}}\hskip.2em
    \subfloat{\includegraphics[width = 0.155\linewidth]{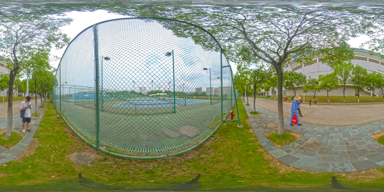}}\hskip.2em
    \subfloat{\includegraphics[width = 0.155\linewidth]{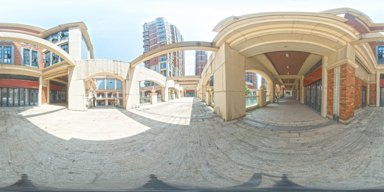}}\hskip.2em
    \subfloat{\includegraphics[width = 0.155\linewidth]{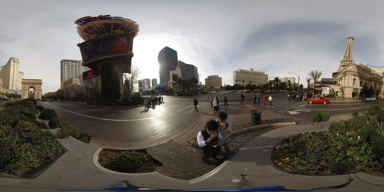}}\hskip.2em

    \subfloat{\includegraphics[width = 0.155\linewidth]{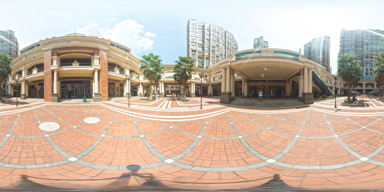}}\hskip.2em
    \subfloat{\includegraphics[width = 0.155\linewidth]{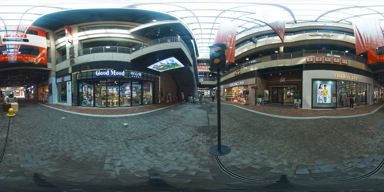}}\hskip.2em
    \subfloat{\includegraphics[width = 0.155\linewidth]{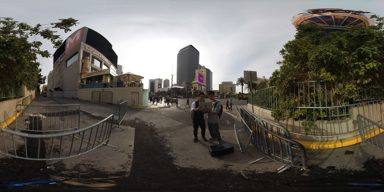}}\hskip.2em
    \subfloat{\includegraphics[width = 0.155\linewidth]{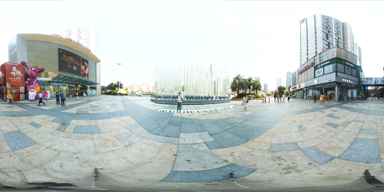}}\hskip.2em
    \subfloat{\includegraphics[width = 0.155\linewidth]{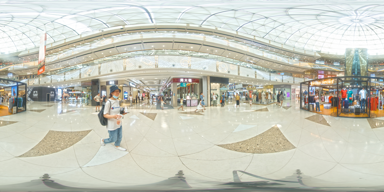}}\hskip.2em
    \subfloat{\includegraphics[width = 0.155\linewidth]{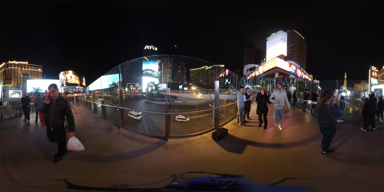}}\hskip.2em
    \caption{Reference images in the proposed database.}
    \label{fig:database_samples}
\end{figure*}

\section{Related Work}
In this section, we first introduce previous psychophysical studies on the perceptual quality of omnidirectional images. We then briefly describe 2D IQA/VQA methods that will serve as building blocks in the proposed computational framework. Last, we review IQA models that are specifically designed for 360{\degree} images.

\subsection{Subjective Quality Assessment of Panoramic Images}
Since the human eye is the ultimate receiver of 360{\degree} images, the most trustworthy way of evaluating visual quality is through psychophysical experiments.
Upenik \emph{et al.} \cite{Upenik2016_subjective} constructed one of the first VR IQA databases to study the impact of compression and projection on the visual quality of panoramas. The absolute category rating was adopted to collect the mean opinion score (MOS) of each image, where a higher MOS means better perceived quality. Additionally, head movement (HM) data was recorded for visual saliency analysis. Duan \emph{et al.} \cite{Duan2018_subjective} built a high-resolution VR IQA database with four distortion types. Apart from the HM data, eye movement (EM) data was recorded for human behavior analysis. They reported that the majority of 2D IQA methods are insufficient to provide accurate quality predictions. Sun \etal \cite{Sun2017_subjective} proposed so far the largest VR IQA database, consisting of $528$ impaired omnidirectional images produced from $16$ references. Huang \etal \cite{Huang2018_subjective} studied the joint effect of spatial resolution and JPEG compression on the perceived quality of 360{\degree} images. Recently, Chen \etal \cite{Chen2020_subjective} conducted a subjective quality assessment of stereoscopic omnidirectional panoramas. The detailed information of these databases is summarized in \autoref{tab:databaes}.

Although several databases included user behavior statistics (\emph{i.e.}, HM/EM data), no timestamp information was found. As a consequence, it is difficult to recover user viewing conditions and behaviors (\emph{e.g.}, the starting point and the scanpath), which are indispensable in omnidirectional IQA. Besides, most previous subjective experiments were carried out on visual materials with global uniform distortions. Little investigation is dedicated to local non-uniform distortions, which may influence user viewing behavior in a substantially different way.

\subsection{Full-Reference Quality Assessment of 2D Images and Videos}
Full-reference IQA and VQA involve developing computational models that are capable of automatically predicting the perceptual quality of images and videos, by comparing to their pristine references. Most full-reference IQA/VQA models are designed for 2D images and videos, among which the mean squared error (MSE) and its derivative peak signal-to-noise ratio (PSNR) are the most widely used. MSE calculates the square differences of pixels between the original and distorted images, and is shown to be poorly correlated with human perception of image quality. Later methods tried to model aspects of the human visual system (HVS) or treated it as a ``black box'' with some holistic assumptions, with the structural similarity (SSIM) index \cite{Zhou2004SSIM_IQA} being the most successful. Recently, there has been a surge of interest in leveraging hierarchical representations of deep neural networks (DNNs) for the design of IQA metrics. Johnson \etal \cite{johnson2016perceptual} used the MSE computed on convolution responses of pre-trained DNNs to guide the optimization of image super-resolution algorithms. Zhang \etal \cite{zhang2018unreasonable} demonstrated the perceptual relevance of deep features pre-trained from a wide range of vision tasks. Ding \etal \cite{Ding2020DISTS_IQA} developed an IQA metric with explicit tolerance to visually similar textures.

Compared with IQA, objective quality assessment of videos is more challenging due to the complex interactions between spatial and temporal distortions. A simple and computationally efficient solution is to compute frame-level quality scores by IQA methods, followed by temporal pooling.
Another type of VQA methods attempted to directly extract spatiotemporal features for quality prediction. Zeng \etal \cite{Zeng2012_3DSSIM} proposed a spatiotemporal SSIM index by treating video signals as 3D volume data.
Kim \etal \cite{Kim2018DeepVQA_VQA} developed a DNN-based full-reference VQA method by incorporating spatiotemporal human visual perception. Xu \etal \cite{Xu2019C3DVQA_VQA} presented a spatiotemporal feature learning framework, where a DNN with 3D convolution kernels was used to learn spatiotemporal distortion thresholds.

\subsection{Objective Quality Assessment of Panoramic Images}
Nearly all quality measures for panoramas adapted existing 2D IQA methods to three formats - 2D plane, sphere, and viewport. Methods \cite{Sun2017WSPSNR_OIQA,Lopes2018WSSIM_OIQA} in the 2D plane tried to compensate for the non-uniform sampling due to sphere-to-plane projection.
Take the equirectangular projection as an example. The local quality measure is weighted by  $\cos(\theta)$, where $\theta$ is the corresponding latitude of the pixel/patch in the spherical domain. In \cite{Zakharchenko2016CPPPSNR_OIQA}, Craster parabolic projection was employed to guarantee uniform sampling density. However, map projections are likely to cause geometric deformations (see Fig.~\ref{fig:sphere2plane}). The second type of methods such as S-PSNR \cite{Yu2015SPSNR_OIQA} and S-SSIM \cite{Chen2018SSIM_OIQA} computed local quality estimates uniformly over the sphere.
Yu \emph{et al.} \cite{Yu2015SPSNR_OIQA} proposed two variants of S-PSNR by deriving importance weightings from statistical distributions of the HM/EM data. The third type of methods\cite{Luz2017SALPSNR_OIQA,Xu2019NCPPSNR_OIQA,Xu2019_viewports} focused on extracting viewports that are highly likely to be explored by viewers.

The above methods \cite{Sun2017WSPSNR_OIQA,Lopes2018WSSIM_OIQA,Zakharchenko2016CPPPSNR_OIQA,Yu2015SPSNR_OIQA,Chen2018SSIM_OIQA,Luz2017SALPSNR_OIQA,Xu2019NCPPSNR_OIQA,Xu2019_viewports} were meaningful attempts to omnidirectional IQA. However, most of them were built on top of traditional 2D IQA models such as PSNR and SSIM, ignoring years of improvements in this field, where more robust and accurate models are available. Previous viewport-based methods only made partial use of the HM/EM data, and did not give a temporal treatment of extracted viewports, making quality assessment ineffective.

\section{Subjective Quality Assessment of 360{\degree} Images}
\label{sec:st}
In this section, we conduct a psychophysical experiment to study the interplay among the VR viewing conditions, the user viewing behaviors, and the perceived quality of omnidirectional images. We first describe the construction of the proposed database, followed by the design of subjective testing and the analysis of human data.

\subsection{Database Construction}
\paragraph{Reference Images} The proposed database\footnote{\url{https://github.com/xiangjieSui/img2video}} contains $36$ pristine-quality images, $24$ of which are captured by us using an insta360 Pro 2 camera, and the remaining $12$ images are downloaded from the Internet\footnote{\url{https://www.insta360.com/cn/product/insta360-pro/\#download-sample}} (carrying a Creative Commons license).
All images have a resolution of $7,860\times3,840$, whose thumbnails are shown in  \autoref{fig:database_samples}.

\begin{figure}
    \centering
    \subfloat{\includegraphics[width = 0.3\linewidth]{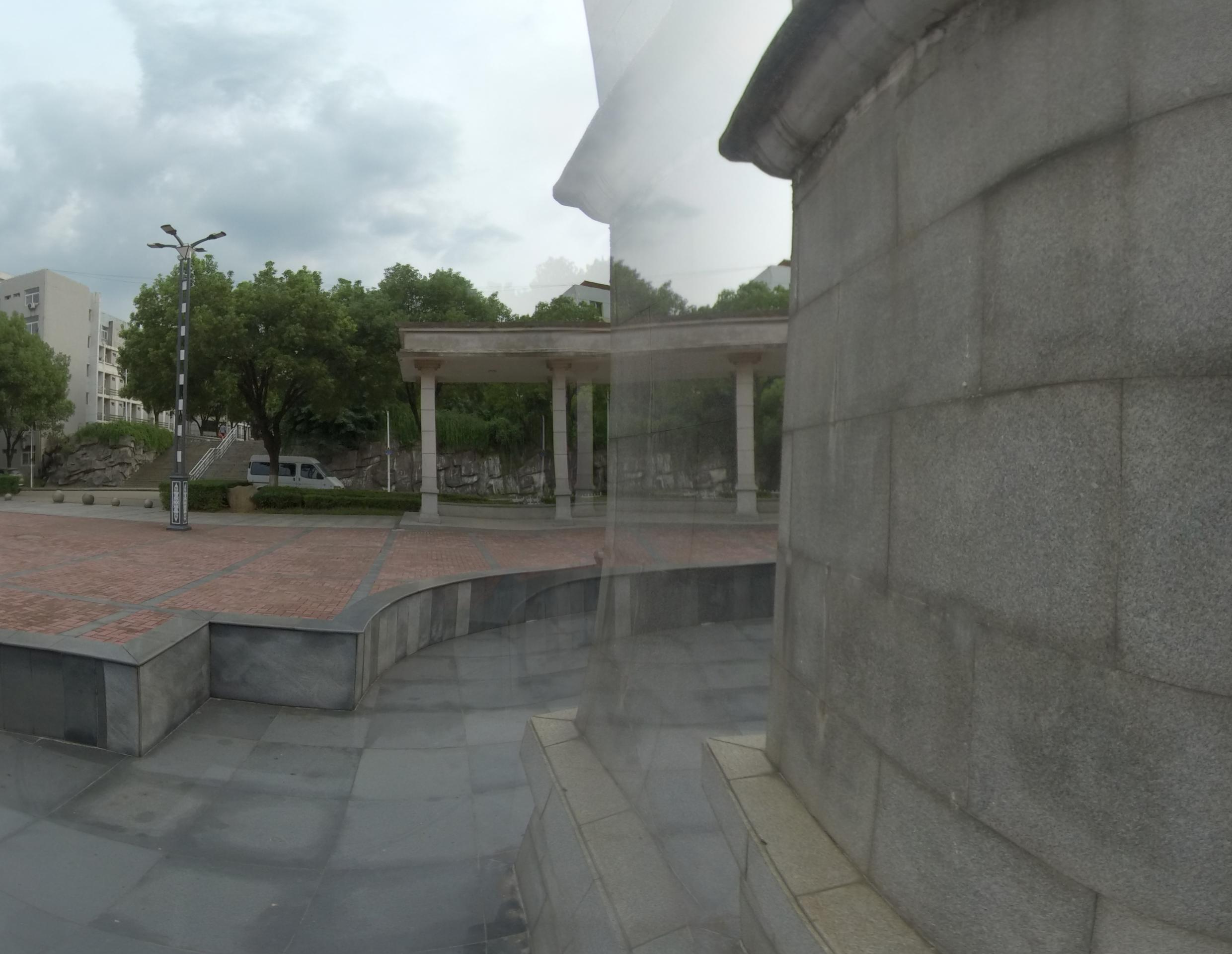}}\hskip.3em
    \subfloat{\includegraphics[width = 0.3\linewidth]{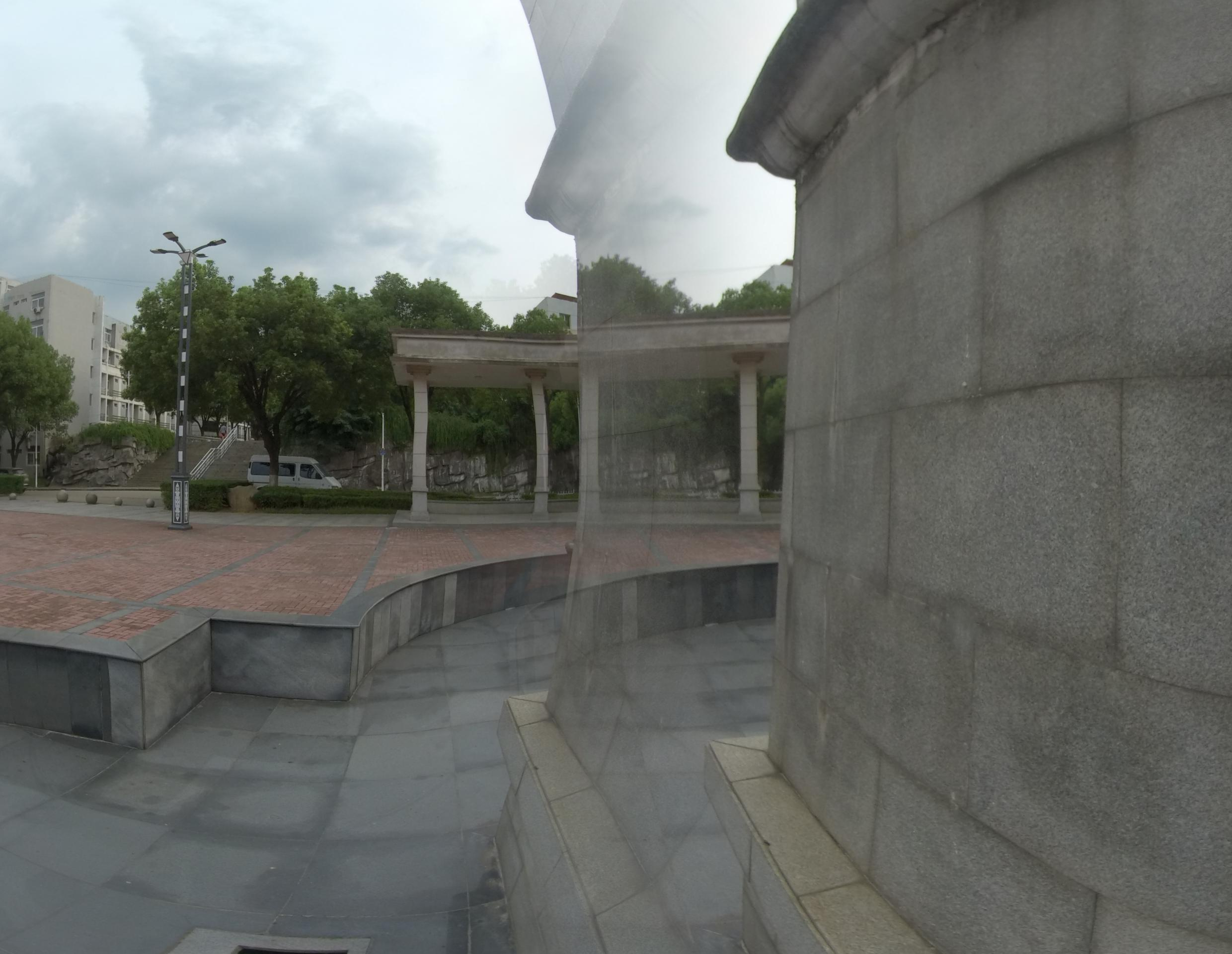}}\hskip.3em
    \subfloat{\includegraphics[width = 0.3\linewidth]{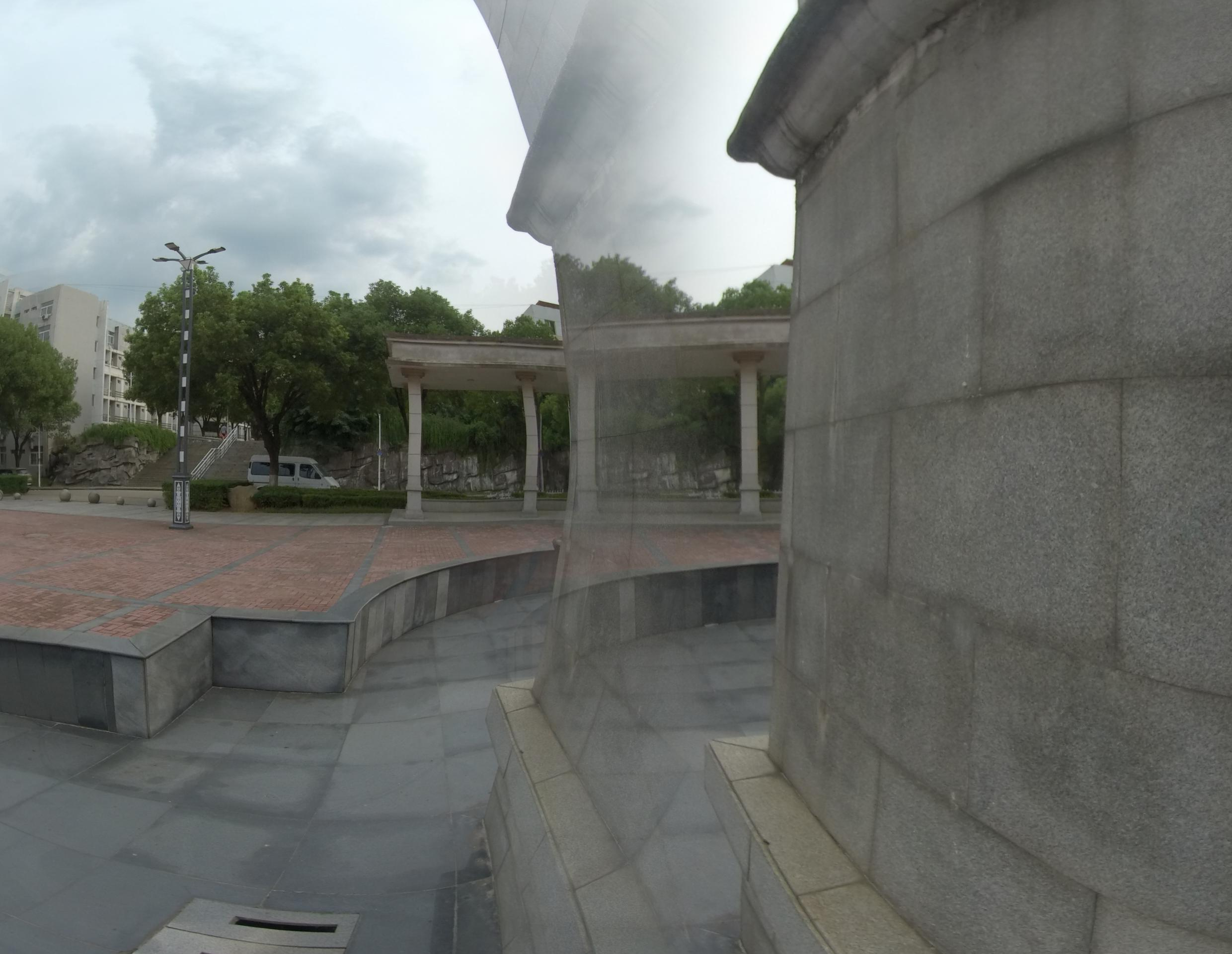}}\hskip.3em
    \caption{Examples of stitching distortions. From left to right, the distortion parameters are set to $0.5$, $0.75$, and $1$, respectively.}
    \label{fig:example_stitching}
\end{figure}

\paragraph{Distorted Images} The driving goal of the subjective test is to investigate how viewing conditions and behaviors affect the perceived quality of omnidirectional images. To this end, two types of distortions are included in this database - stitching distortion and H.265 compression. Stitching artifacts occur when images of different views are not properly aligned when stitching. In our database, the popular software toolbox Nuke\footnote{\url{https://www.foundry.com/products/nuke}} is employed to generate stitching distortions. We first import the raw images (\ie, fisheye images in the format of .dng) into Nuke, and adjust the distortion parameter of one
lens to control the level of radial distortions. We choose the stitching distortion parameters from $\{ 0.5, 0.75,1\}$ with higher values implying severer distortions (see \autoref{fig:example_stitching}).
We use the FFmpeg libx265 encoder to generate compression artifacts by setting the quantization parameters to $\{38, 44, 50\}$ with higher values indicating increased distortions.
We apply stitching distortions to the $24$ images captured by us, and compress the $12$ images downloaded from the Internet. In summary, the proposed database consists of $72$ images, half of which are distorted.

\paragraph{Viewing Conditions} We consider two viewing conditions: the starting point and the exploration time. Two starting points are chosen for each image: one for a salient region (\eg, due to the localized stitching distortion) and the other on the opposite side of the former (\eg, where the stitching distortion is not visible within the initial viewport). Similarly, two time periods, \ie, $5$ and $15$ seconds, are set for each image. Totally, there are $2\times 2=4$ viewing conditions.

\begin{figure}[!t]
  \centering
  \subfloat[]{\includegraphics[width=\linewidth]{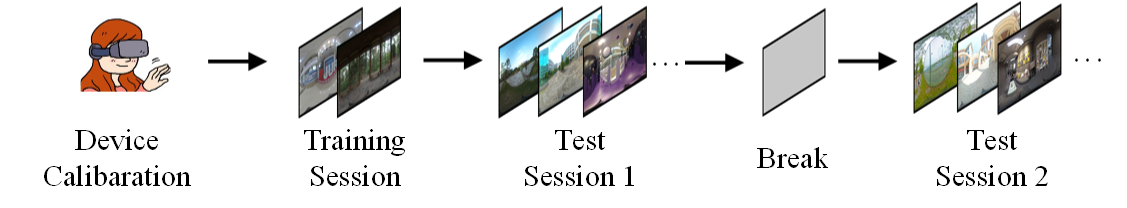}}\\
  \subfloat[]{\includegraphics[width=\linewidth]{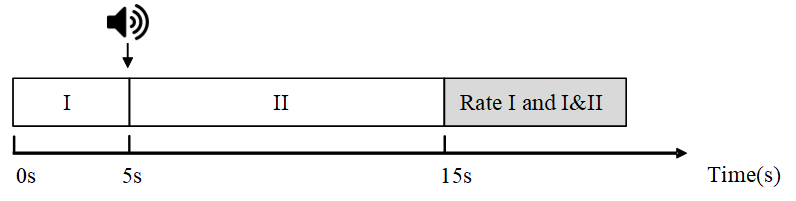}}
  \caption{The design of our psychophysical experiment. (a) Overall experimental procedure, which consists of one training session, two test sessions with a $10$-minute break in between. Each test session contains $18$ distorted omnidirectional images. (b) Phase I: the first $5$ seconds of viewing and Phase II: the last $10$ seconds of viewing, separated by a voice prompt. The subjects need to give two scores to indicate their viewing experience in Phase I, and both Phase I and Phase II.}
  \label{fig:subjective_test_procedure}
\end{figure}

\subsection{Psychophysical Experiment Design}
\label{sec:st_design}
We employ the single stimulus continuous quality evaluation method described in the ITU-R BT 500.13 recommendation \cite{series2012methodology} to gather human data. Subjects are asked to rate the quality of an omnidirectional image on a continuous scale of $[1, 5]$, labeled by five quality levels (``bad'', ``poor'', ``fair'', ``good'', and ``excellent''). The images are displayed in a random order using an HTC Vive VR HMD, which provides a field of view (FoV) of 110{\degree}. EM and HM data are collected by a built-in Tobii Pro eye tracking system with a fixed sampling frequency of $20$ Hz. Image playback is supported by a high-performance server with an AMD Ryzen 9 3950X $16$-Core CPU, a 128 GB RAM, and an NVIDIA GeForce RTX 2080 Ti GPU. The user interface is built by the Unity Game Engine.

We invite $22$ subjects to participate in the psychophysical study. They are divided into two groups, and are asked to view 360{\degree} images from two different sets of starting points. We conduct two test sessions with a 10-minute break in between to minimize the effect of fatigue and discomfort (see \autoref{fig:subjective_test_procedure} (a)). A training session is included to familiarize the subjects with the rating procedure and to exclude those who feel discomfort exploring VR environments.  We also design a rating strategy to collect quality scores with different exploration periods (see \autoref{fig:subjective_test_procedure} (b)). A voice prompt is played when the subjects have viewed a 360{\degree} image for $5$ seconds to remind them of giving a quality score based on their viewing experience so far. When the subjects finish viewing the image within $15$ seconds, they need to give another quality score according to their overall viewing experience. It is worth noting that each image is viewed only once by one subject to ensure that user data is collected without prior knowledge of the scene. To sum up, in the proposed database, each image is associated with five tags, including the starting point, the exploration time, the distortion type, the distortion level, and the MOS.

\subsection{Psychophysical Data Analysis}
\begin{figure}
  \centering
   \subfloat[]{\includegraphics[width=0.45\linewidth]{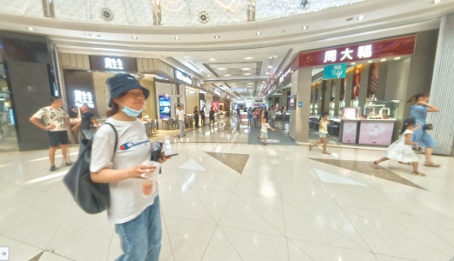}}\hskip.4em
   \subfloat[]{\includegraphics[width=0.45\linewidth]{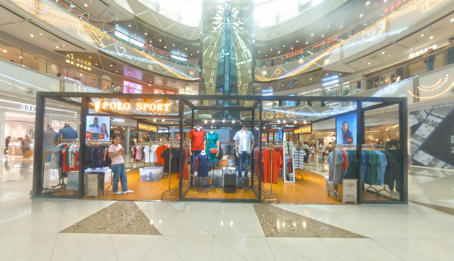}}
  \caption{Consistency of user viewing behaviors under  different starting points. The exploration time is fixed to $5$ seconds. (a) Initial viewport that contains a passerby, attracts human visual attention, leading to a higher PLCC of $0.935$. (b) Initial viewport that exhibits symmetrical image structures with no eye-catching event,  results in a much lower PLCC of $0.187$.}
  \label{fig:subject_test_st}
\end{figure}

\begin{figure}[!t]
  \centering
  \subfloat[]{\includegraphics[width=0.48\linewidth]{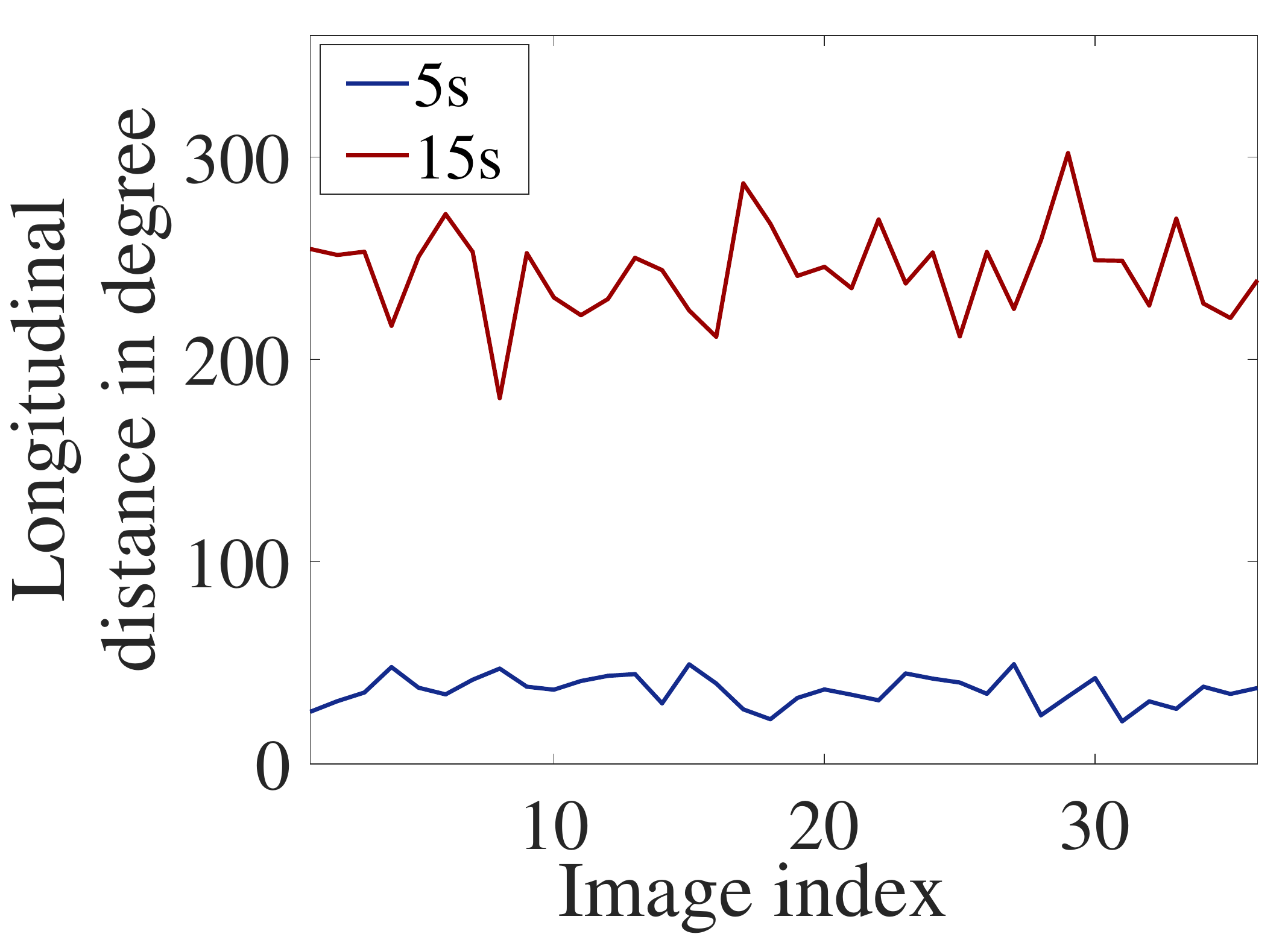}}\hskip.3em
  \subfloat[]{\includegraphics[width=0.48\linewidth]{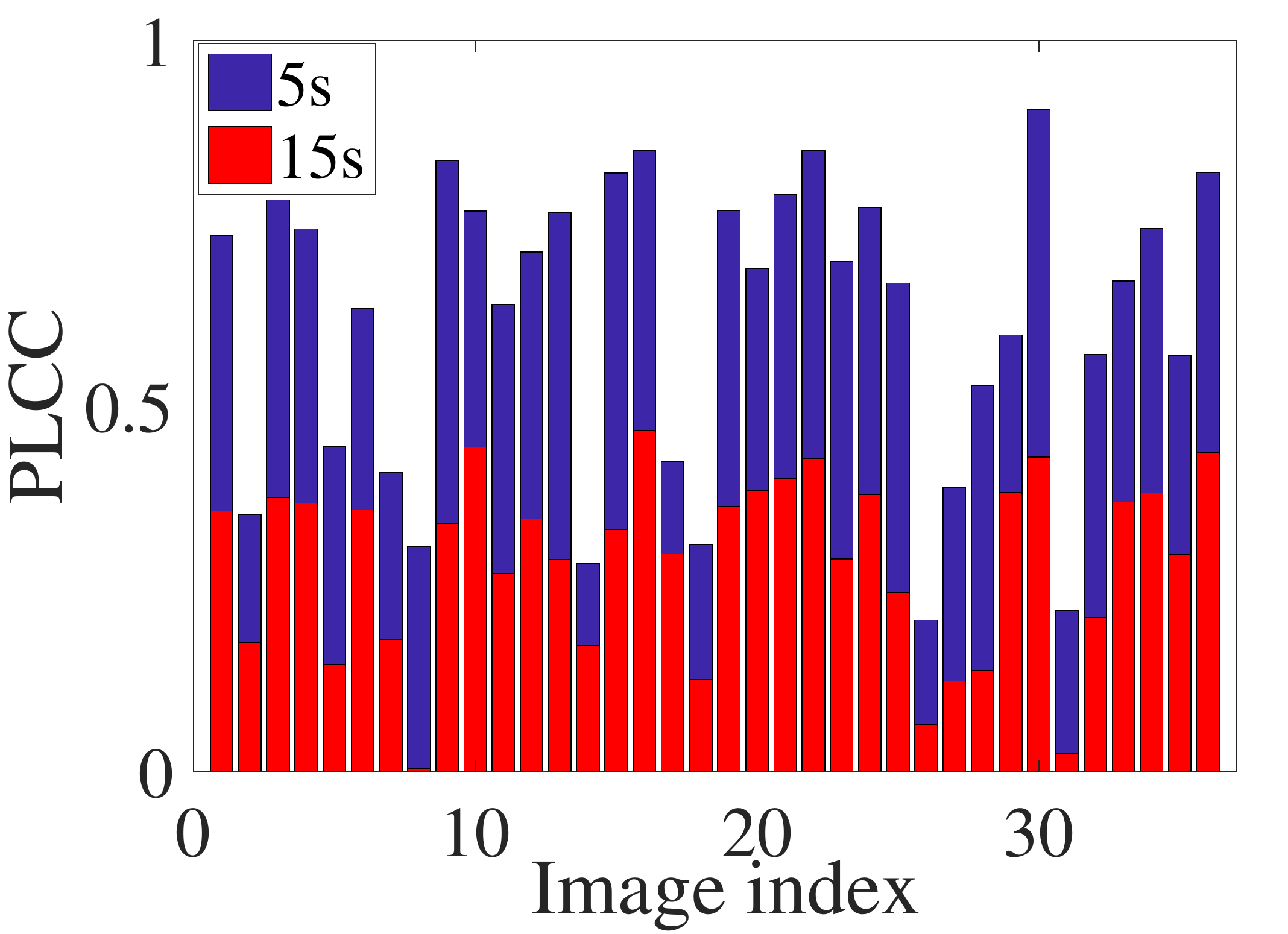}}
  \caption{(a) Farthest longitudinal distance to the starting point averaged across viewers for $5$ and $15$  seconds of exploration. (b) Consistency between scanpaths from different users in terms of PLCC.}
  \label{fig:subject_test_time}
\end{figure}

\begin{figure*}[htb]
    \centering
    \subfloat[$5$ seconds, stitching]{\includegraphics[width=0.24\textwidth]{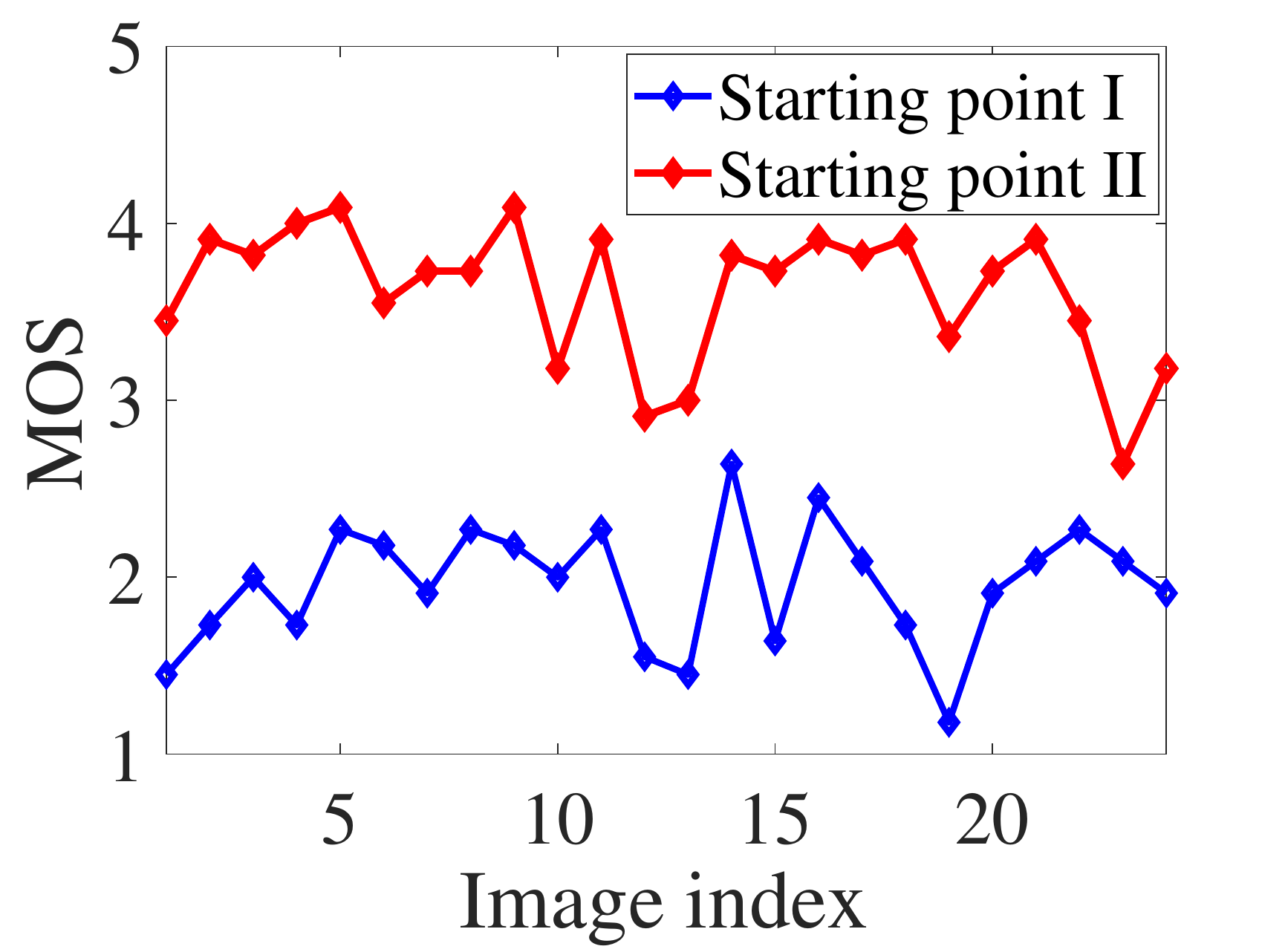}}\hskip.2em
    \subfloat[$15$ seconds, stitching]{\includegraphics[width=0.24\textwidth]{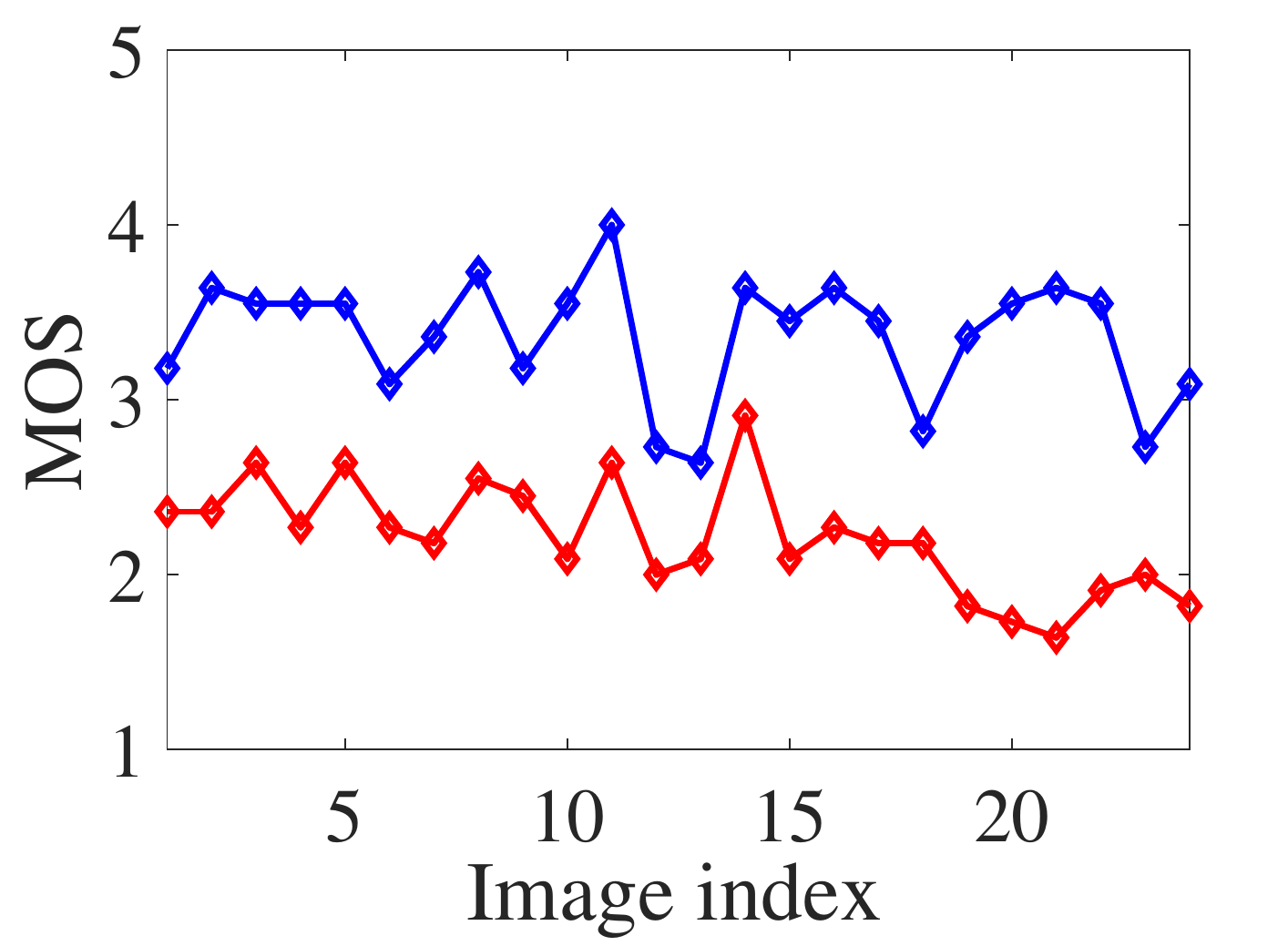}}\hskip.2em
     \subfloat[$5$ seconds, compression]{\includegraphics[width=0.24\textwidth]{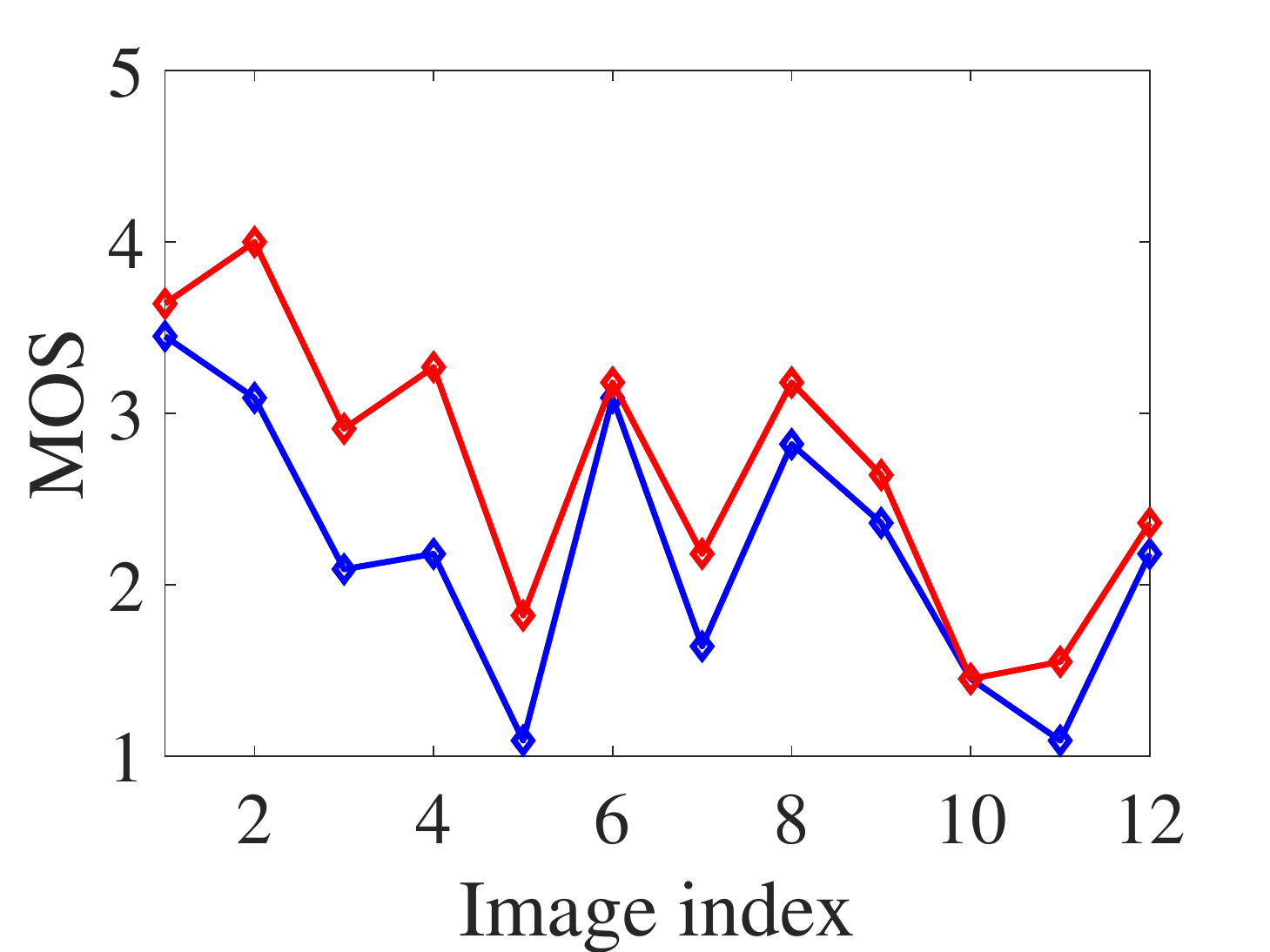}}\hskip.2em
    \subfloat[$15$ seconds, compression]{\includegraphics[width=0.24\textwidth]{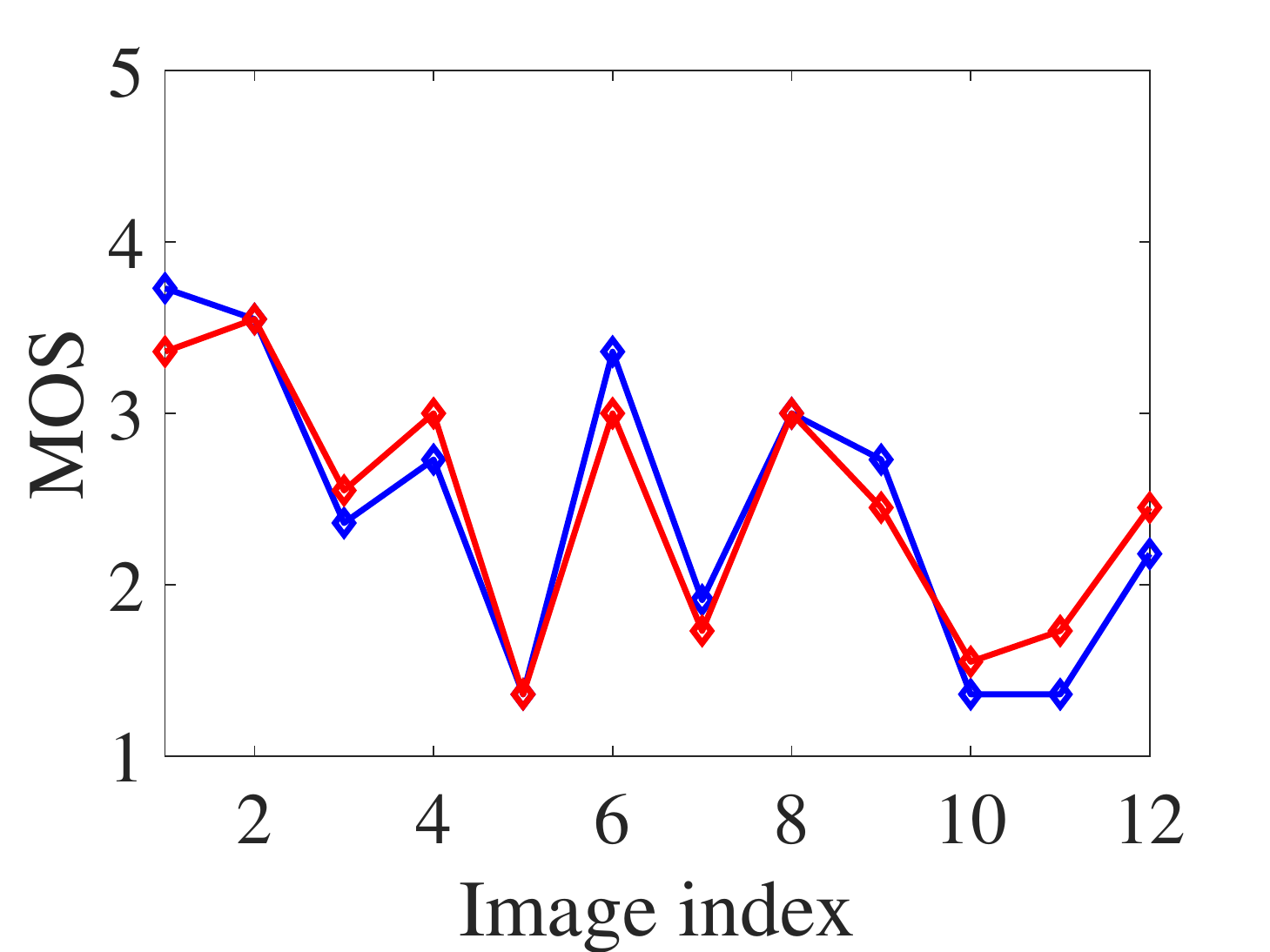}}\\
    \subfloat[Starting point I, stitching]{\includegraphics[width=0.24\textwidth]{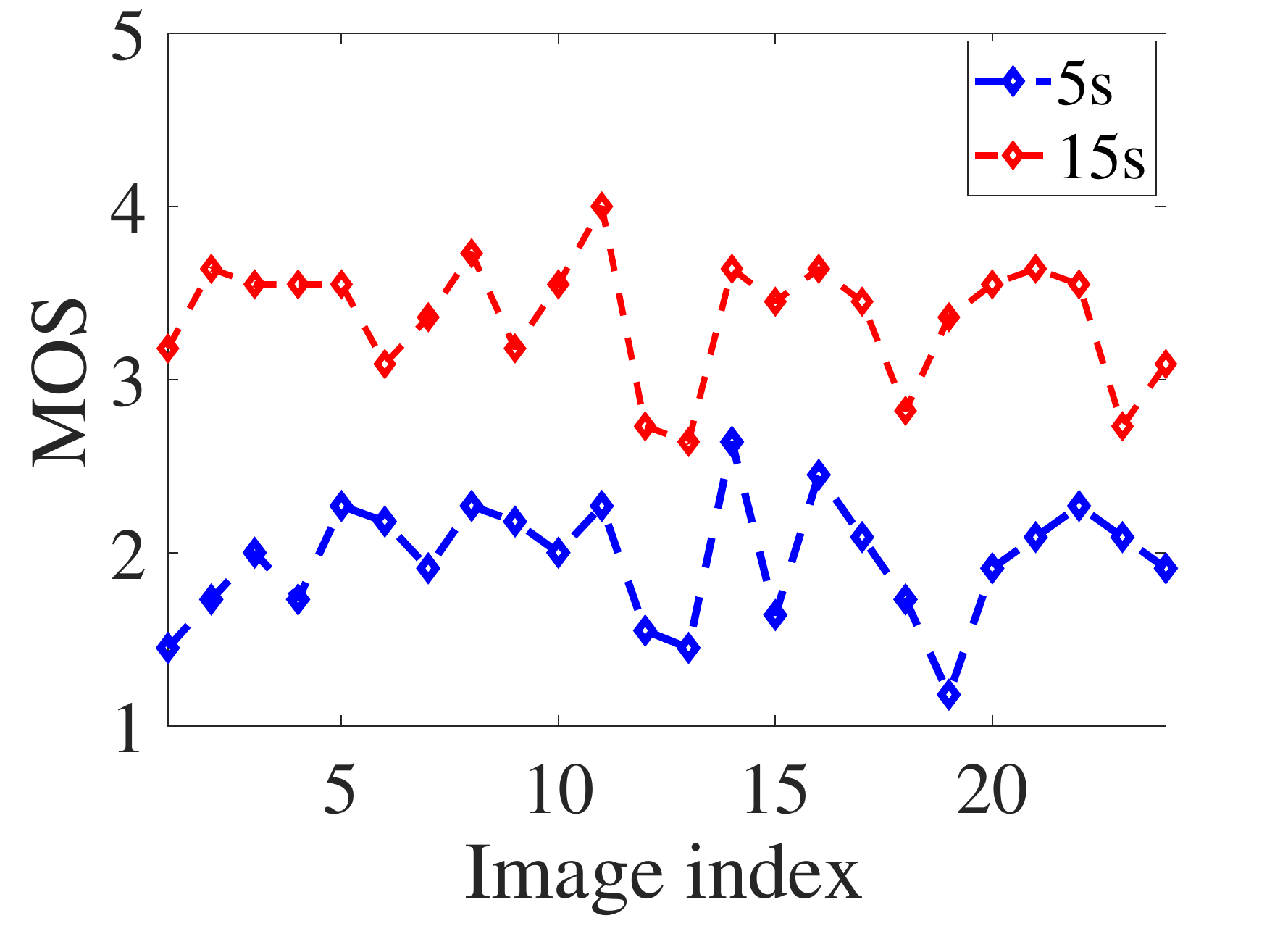}}\hskip.2em
    \subfloat[Starting point II, stitching]{\includegraphics[width=0.24\textwidth]{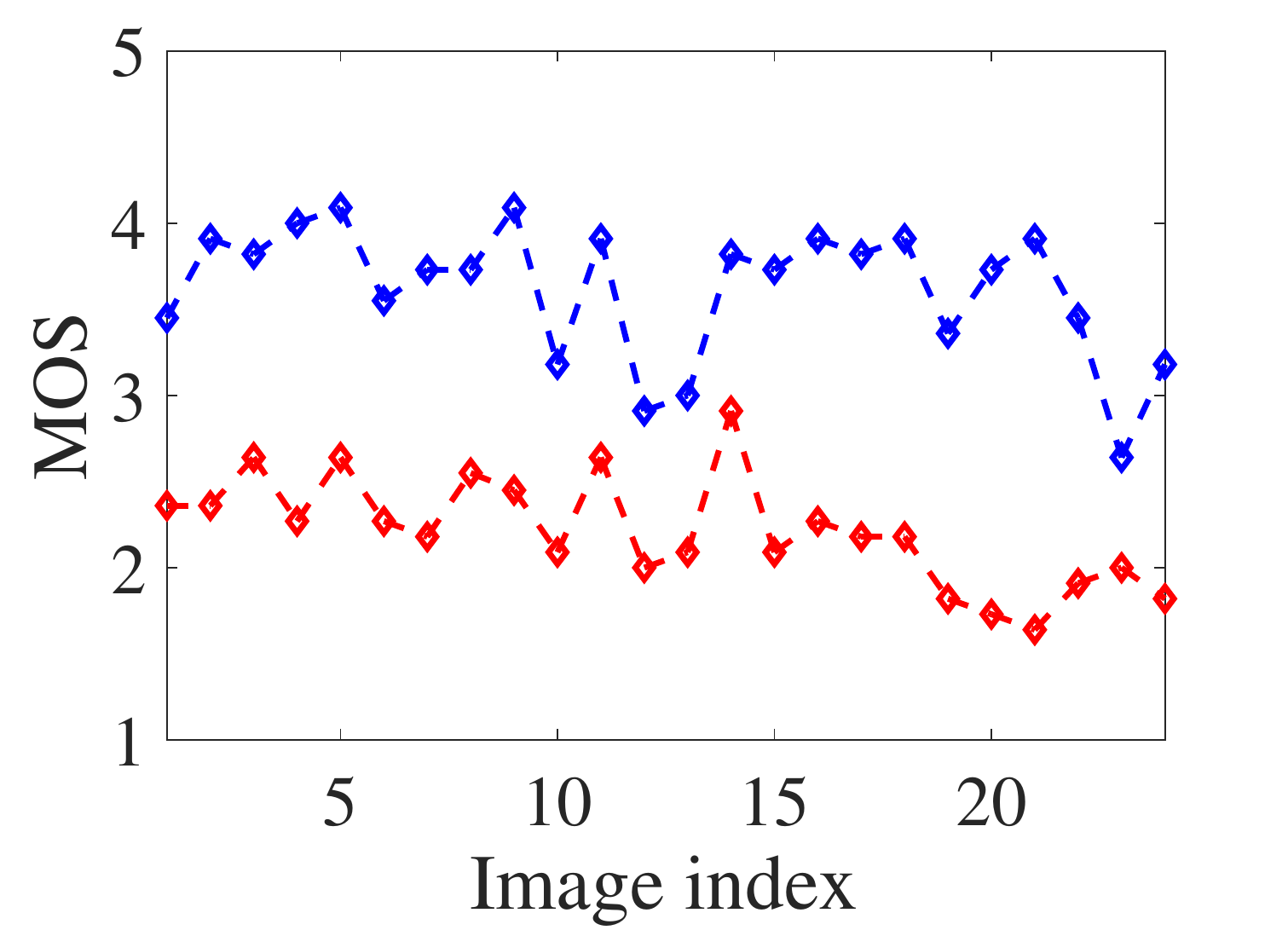}}\hskip.2em
    \subfloat[Starting point I, compression]{\includegraphics[width=0.24\textwidth]{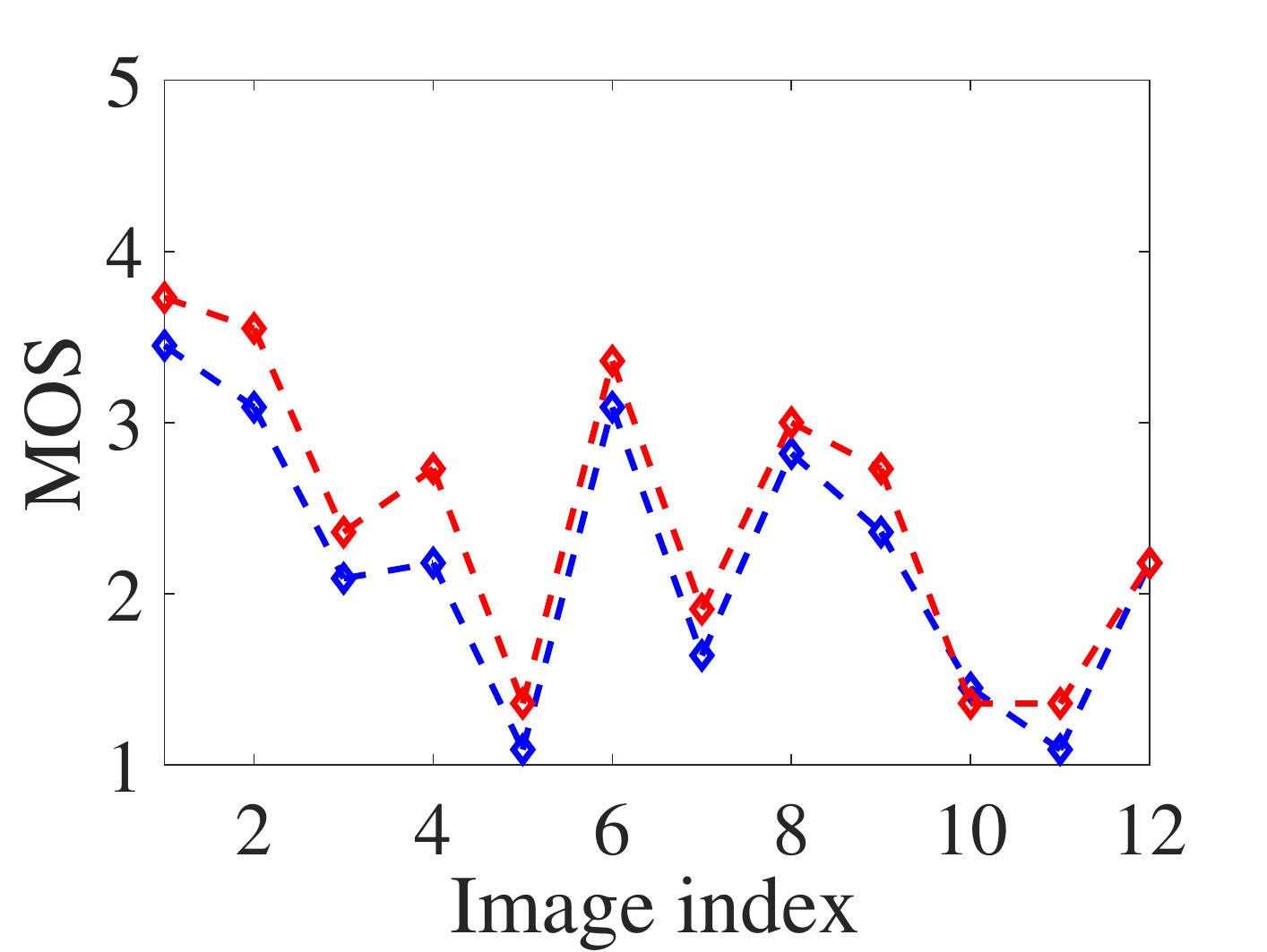}}\hskip.2em
    \subfloat[Starting point II, compression]{\includegraphics[width=0.24\textwidth]{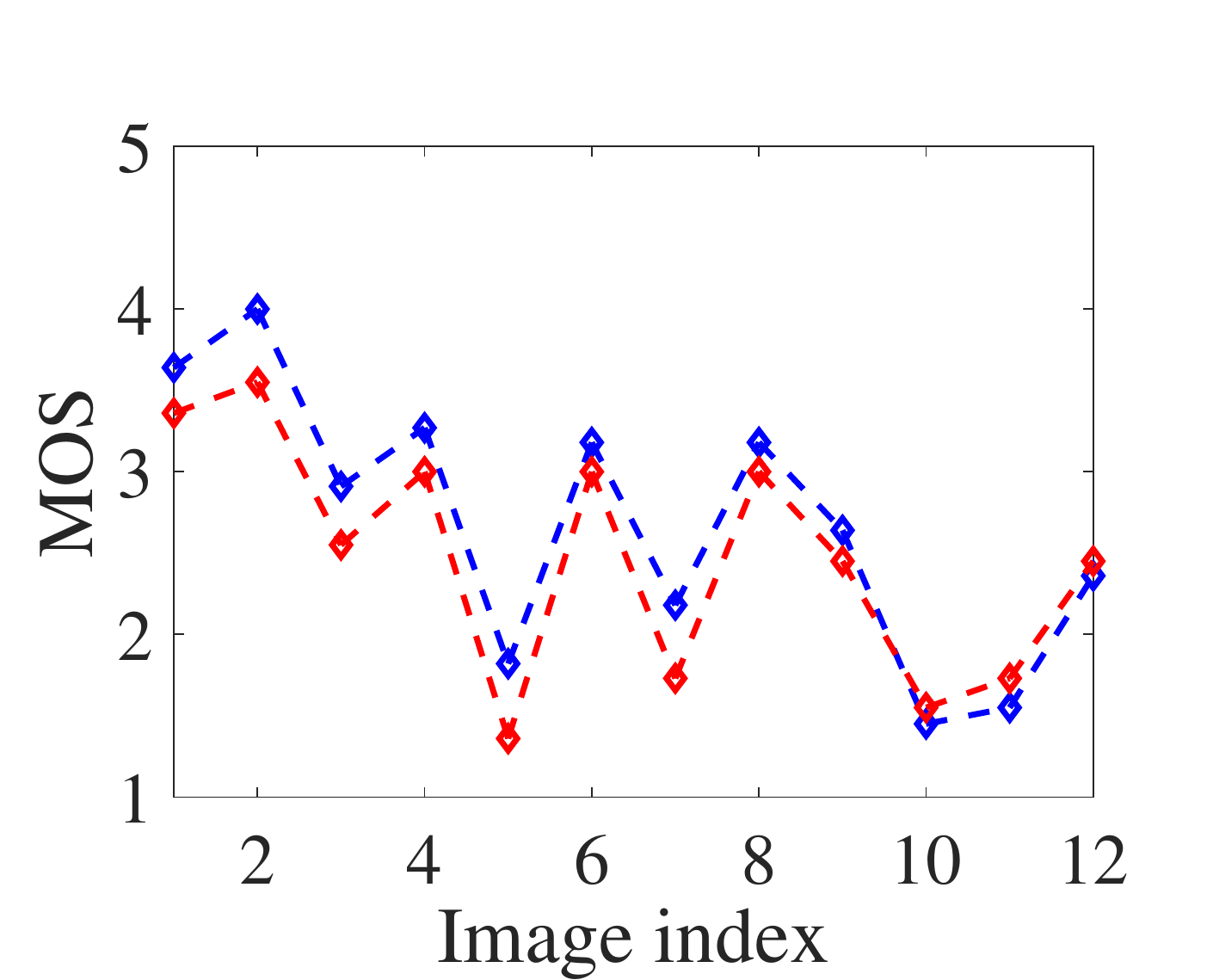}}
    \caption{MOSs of 360{\degree} images in the proposed database under different viewing conditions.}
    \label{fig:MOS_viewing_conditions}
\end{figure*}

\subsubsection{Do Viewing Conditions Affect Viewing Behaviors?}
We arrange the HM data gathered from each user in a matrix, where the row dimension records the latitude and the longitude, and the column dimension records the time instance. For each image, we compute the average Pearson linear correlation coefficient (PLCC) of the HM matrices across every pair of users under the same viewing condition, as an indication of user behavior consistency.

\begin{table}[h]
  \centering
  \small
  \begin{tabular}{l|l|l|l|l|l}
    \toprule
    \textit{Source of variation} & \textbf{$SS$} & \textbf{$d.f.$} & \textbf{$MS$} & \textbf{$F$} & \textbf{$p$} \\ \hline
    Starting point & $2.76$ & $1$ & $2.76$ & $9.60$ & $\approx 0$\\ \hline
    Exploration time & $0$ & $1$ & $0$ & $0.01$ & $0.91$\\ \hline
    Distortion type & $3.49$ & $1$ & $3.49$ & $12.14$ & $\approx 0$\\ \hline
    \tabincell{l}{Starting point $\times$\\Exploration time} & $19.06$ & $1$ & $19.10$ & $66.28$ & $\approx 0$\\ \hline
    \tabincell{l}{Starting point $\times$\\Distortion type} & $0.08$ & $1$ & $0.08$ & $0.29$ & $0.59$\\ \hline
    \tabincell{l}{Exploration time $\times$\\Distortion type} & $0$ & $1$ & $0$ & $0.01$ & $0.90$\\ \hline
    \tabincell{l}{Starting point $\times$\\Exploration time $\times$\\Distortion type} & $9.45$ & $1$ & $9.45$ & $32.86$ & $\approx 0$\\ \hline
    Residual & $39.11$ & $136$ & $0.29$& & \\
    \hline
    Total & $88.31$ & $143$ & & & \\
    \bottomrule
  \end{tabular}
  \caption{The results of multi-factorial ANOVA test. $SS$: sum of squares. $d.f.$: degrees of freedom. $MS$: mean square. $F$: $F$ value. $p$: $p$-value for the null hypothesis. }\label{table:ANOVA}
\end{table}

\paragraph{How Does the Starting Point Affect  Viewing Behaviors?}
It is natural to infer that the starting point has an important influence on the gaze scanpath. For example, salient targets in the initial viewport may attract attention, leading to more consistent gaze scanpaths at least for a short time period (see Fig. \ref{fig:subject_test_st} (a)). If the image structures of the initial viewport are symmetric, the gaze scanpaths tend to be random, as the users have equal probability to start exploring the scene from left or right (see Fig. \ref{fig:subject_test_st} (b)).

\paragraph{How Does the Exploration Time Affect Viewing Behaviors?}
The exploration time directly affects how many viewports can be observed by the users. \autoref{fig:subject_test_time} (a) shows the farthest longitudinal distances to the starting point averaged across users for $5$ and $15$ seconds of exploration. When a 360{\degree} image is displayed, the subjects usually take some time to adapt to the new scene, and $5$ seconds may not be sufficient for them to fully explore the scene. In contrast, most users are able to finish viewing the image within $15$ seconds.
Fig. \ref{fig:subject_test_time} shows the PLCC results for different exploration time, where we observe that the correlation between scanpaths from different viewers decreases substantially over time.

\subsubsection{Do Viewing Conditions Affect Perceived Quality?}

Assuming a reasonable gaze speed, a short exploration time means fewer observed viewports, which may highlight the starting point in the quality assessment process. By contrast, with a long exploration time, the viewers are more likely to be influenced by viewports close to the end of viewing due to the recency effect~\cite{Hands2001Recency}. These hypotheses have been validated in our psychophysical experiment. We show the MOSs of the images under different viewing conditions in \autoref{fig:MOS_viewing_conditions}, and have two important findings:
\begin{itemize}
\item Both the starting point and the exploration time have a noticeable impact on the perceived quality of 360{\degree} images with \textit{localized distortions} (\ie, stitching distortions). However, they seem to have little effect on H.265 compressed images with global uniform distortions. Therefore, we have identified the distortion type as a determining factor on how the viewers respond to different viewing conditions.

\item The \textit{recency effect} is clearly observed when the users explore locally distorted omnidirectional images (see Figs. \ref{fig:MOS_viewing_conditions} (e) and (f)).
From Starting Point I where stitching distortions appear in initial viewports, the users usually give low quality scores after $5$ seconds of viewing (\ie, in Phase I). However, if the viewers are allowed to explore the panoramic scene for $15$ seconds, the quality ratings are considerably higher.
On the contrary, from Starting Point II, where stitching distortions are at the opposite side of the initial viewports, the users would probably see such localized distortions in Phase II. As a result, they tend to give high and low quality scores for $5$ and $15$ seconds of viewing, respectively.
\end{itemize}

To test the significance of the starting point, the exploration time, and the distortion type on influencing the perceived quality of omnidirectional images, we apply the multi-factorial analysis of variance (ANOVA) \cite{Tabachnick2007}, which considers all factors at once. The results are summarized in \autoref{table:ANOVA}, from which we identify two significant \textit{individual effects}: the starting point and the distortion type (whose $p$-values are below the threshold of $0.05$). The exploration time cannot alone explain the variability in perceived quality (with $p$-value $=0.91$). A statistically significant \textit{interplay effect} is also identified, implying that the perceived quality depends on the combination of the starting point and the exploration time. These results show that the viewing conditions have a significant impact on the perceived quality. Nevertheless, it is reasonable to assume that the perceived quality depends on the combination of the starting point, the exploration time, and the distortion type, despite that the corresponding $F$ value is not the largest in \autoref{table:ANOVA}.

\section{Objective Quality Assessment of 360{\degree} Images}
\begin{figure*}[tb]
 \centering
 \includegraphics[width=0.85\textwidth]{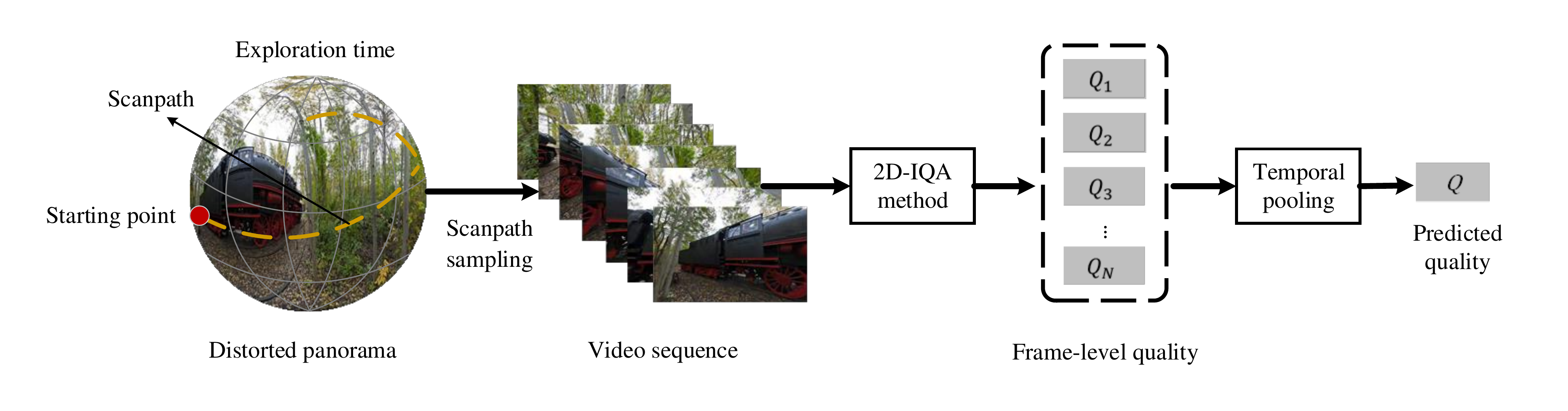}
 \caption{Proposed computational framework for omnidirectional IQA. The processing of the reference panorama has been omitted for simplicity.}
 \label{fig:framework}
\end{figure*}
In this section, we describe a general computational framework for omnidirectional IQA, where user viewing conditions and behaviors are incorporated naturally by treating omnidirectional images as moving camera videos, as shown in \autoref{fig:framework}.

\subsection{Input Data}
The inputs to our computational model consist of 1) a pair of reference and distorted panoramas, 2) a starting point and an exploration time as two types of viewing conditions, and 3) scanpaths from different users as one type of  viewing behavior:

\begin{itemize}
\item \textbf{The reference and distorted panoramas}, $X$ and $Y$, are usually in the form of 2D equirectangular projections, which are to be transformed to spherical representations for viewport extraction.
    \item \textbf{The starting point}, $P_0 = (\phi_0,\theta_0)$, specifies the longitude and the latitude, at which the initial viewport is centered for a viewer to start exploring the virtual scene.

    \item \textbf{The exploration time}, $T$, records how long it takes for a viewer to explore the distorted panorama.

    \item \textbf{The scanpath}, $P(t): \mathbb{R}\mapsto \mathbb{R}^2$, describes a 2D gaze trajectory when exploring the visual field~\cite{Noton1971Scanpath_Intro}. It takes a time instant $t\in[0, T]$ as input, and produces a 2D spherical coordinate $(\phi, \theta)$, where $P(0) = (0,0)$. The viewport at a specific time instant $t$ can be extracted at $P(t) + P_0$.
\end{itemize}

\subsection{Omnidirectional Image-to-Video Conversion}
Given the viewing behavior of a user, we convert a panorama into a video sequence, which contains only global motion as if the underlying static scene were captured by a moving camera. This is achieved by sampling a sequence of rectilinear projections of viewports of the panorama along the scanpath \cite{Ye2017JVET_360Lib}, with a predefined sampling rate.
Specifically, given the current sample point $P(t)+P_0$ as the center, we first set the FoV to $[-\pi/6,\ \pi/6]$ along both longitude and latitude directions, inspired by the theory of near peripheral vision \cite{Besharse2011retina}. This specifies 3D Cartesian coordinates of the square viewport, which are assumed to be perpendicular to the Z-axis for convenience. The corresponding pixel values can be retrieved by projecting 3D Cartesian points onto the unit sphere and then onto the 2D plane. Bilinear interpolation is used as the optional resampling filter. We then choose a sampling rate
\begin{align}\label{eq:sr}
R = \frac{1}{s_1}\times R_{\mathrm{et}},
\end{align}
where $R_{\mathrm{et}}$ is the maximum sampling rate constrained by the eye tracker and $s_1\ge1$ is a stride parameter. The resulting moving camera video has  a total of $N=R\times T$ frames.

\subsection{Omnidirectional Image Quality Prediction}
Generally, any existing VQA model could be adopted at this stage to evaluate the perceived quality of 360{\degree} images. Here we follow a two-stage approach: frame-level quality estimation followed by temporal pooling. For the $i$-th viewer, where $i\in\{1,2\ldots,M\}$, we denote the $j$-th frames of the reference and distorted videos by $X_{ij}$ and $Y_{ij}$, respectively. The frame-level quality can then be computed by
\begin{align}
Q_{ij} = D\left(X_{ij}, Y_{ij}\right),
\end{align}
where $D$ denotes a full-reference IQA model.  The global quality $Q_i$ as perceived by the $i$-th user  can be computed by fusing frame-level quality scores:
\begin{align}
    Q_{i} = F(Q_{i1},\ldots, Q_{iN}),
\end{align}
where $F$ is a temporal pooling strategy that may model aspects of the memory effect of the human brain. Similar as computing the MOS, we average quality estimates across all viewers to obtain the final quality score of the distorted panorama:
\begin{align}
Q = \frac{1}{M} \sum_{i=1}^{M}Q_i.
\end{align}

\subsection{Specific Omnidirectional IQA Models}\label{subsec:ins}
We construct several specific examples of omnidirectional IQA measures within the proposed computational framework. First, we need to specify 2D IQA models for computing frame-level quality. The main selection criterion is that the model should correlate well with human perception of image quality, in terms of \textit{benchmarking} as well as \textit{optimizing} image processing algorithms. In this paper, we select five full-reference image quality models:
\begin{itemize}
    \item PSNR, the Peak Signal-to-Noise Ratio, is built on top of the MSE by incorporating the maximum power of a signal. Arguably PSNR (or MSE) is the most widely used IQA measure, and enjoys a number of desirable properties for optimization purposes.
    \item SSIM~\cite{Zhou2004SSIM_IQA}, the Structural SIMilarity index, assumes that the HVS is highly adapted to extract local image structures of the visual field. Thus, a measure of structural information loss may provide a good approximation to perceived quality degradation. Over the years, SSIM and its multi-scale extension~\cite{Wang2003MSSSIM} have been regarded as standard ``perceptual'' metrics to guide the optimization of methods for image denoising~\cite{Channappayya2008denoising}, image compression~\cite{Ball2018compression}, image synthesis~\cite{Snell2017synthesis}, and video coding~\cite{Wang2012videocoding}.
    \item VIF~\cite{Sheikh2005VIF_IQA}, the Visual Information Fidelity measure, offers an information theoretical perspective of IQA, and uses the mutual information~\cite{Sheikh2005inforFidelity} to quantify the amount of information preserved in the distorted image. Its industrial implementation - VMAF~\cite{VAMF2016techBlog} has been successfully applied to adjust
   the parameter settings in video engineering.
    \item NLPD \cite{Laparra2017NLPD_IQA}, the Normalized Laplacian Pyramid Distance, is based on a multi-scale nonlinear representation that models the operations in early stages of the HVS. NLPD has been used to optimize tone mapping algorithms, where the input image has a much higher dynamic range than that of the output image \cite{Laparra2017NLPD_IQA}.
    \item DISTS~\cite{Ding2020DISTS_IQA}, the Deep Image Structure and Texture Similarity metric, uses a DNN to construct an injective and perceptual transform, and makes SSIM-motivated quality measurements in the transform domain. DISTS is robust to texture substitution and mild geometric transformation. In a recent comparison of IQA models for optimization of image processing systems~\cite{Ding2020optimization}, DISTS outperforms ten competing models in blind image deblurring, single image super-resolution, and lossy image compression.
\end{itemize}

We adopt the temporal hysteresis model~\cite{Seshadrinathan2011TH_VQA} as the default pooling strategy. Specifically, to mimic users' intolerance  to poor quality events and reluctant reaction  to quality improvement events, a memory component is defined at each video frame:
\begin{align}
Q^m_j=
 \begin{cases}
 Q_{1} & \mbox{if } j = 1 \\
 \min\left\{Q^m_{\max\{1, j-K\}},\ldots,Q^m_{j-2},Q^m_{j-1}\right\}& \mbox{otherwise}
\end{cases}
\end{align}
where we omit the user index $i$ in the subscript to make the notation uncluttered. $K$ is a parameter related to the duration of memory \cite{Seshadrinathan2011TH_VQA}.

The temporal hysteresis pooling also accounts for the fact that humans react sharply to quality degradation events by defining a current quality component at each video frame:
\begin{align}
Q^c_j= \sum_{k=j}^{\min\{j+K,N\}}w_{k}Q^s_{k},
\end{align}
and
\begin{align}
\left\{Q^s_k\right\} = \mathrm{sort}\left(\left\{Q_{k}\right\}_{k=j}^{\min\{j+K,N\}}\right),
\end{align}
where $\mathrm{sort}()$ sorts $\{Q_{j}, \ldots, Q_{\min\{j+K,N\}}\}$ in ascending order, resulting in $\{Q^s_{j}, \ldots, Q^s_{\min\{j+K,N\}}\}$. 
$w$  is a normalized weighting vector specified by the descending half of a Gaussian function. The adjusted time-varying quality score of $Y_{j}$ is computed by linearly combining the memory and current components:
\begin{align}
Q^a_{j} = \alpha Q^m_{j} + (1-\alpha)Q^c_{j},
\end{align}
where $\alpha$ is a parameter to trade off the two terms. The global quality is obtained by averaging the quality scores of all frames:
\begin{align}
Q = \frac{1}{N} \sum_{j=1}^{N}Q^a_{j}.
\end{align}

\begin{table}[tb]
  \renewcommand\arraystretch{1.4}
  \scriptsize%
  \centering%
  \begin{tabular}{ c| p{5.5cm}}
      \toprule
   & Value \\  \hline
  $P_0$ & \{$(-\frac{\pi}{2},0)$,\ $(0,0)$,\ $(\frac{\pi}{2},0)$,\ $(\pi,0)$\} \\   \hline
  $T$ & $15$ seconds \\   \hline
  $P(t)$ & $\phi = \begin{cases}
  -v t, 0\le t \leq \frac{T}{4} \\[4pt]
 -\frac{\pi}{2} + v (t-\frac{T}{4}), \frac{T}{4} < t \leq \frac{3T}{4}, \\[4pt]
\frac{\pi}{2} - v (t-\frac{3T}{4}),  \frac{3T}{4} < t \leq T
\end{cases} \theta=0$
  \\\bottomrule
  \end{tabular}
  \caption{Default viewing conditions and behaviors for panoramic image-to-video conversion. $(\phi, \theta)$ are the longitude and the latitude, and $v$ is the gaze velocity.}
  \label{tab:default_setting}
\end{table}

\begin{table*}[!htb]
    \centering
    \small
    \renewcommand\tabcolsep{5pt}
    \begin{tabular}{r|rrrrrr|rrrrrrrrrr }
    \toprule
    & \multicolumn{6}{c|}{ Proposed database} & \multicolumn{10}{c}{OIQA database} \\ \hline
         & \multicolumn{2}{c}{ST} & \multicolumn{2}{c}{H.265} & \multicolumn{2}{c|}{Overall} & \multicolumn{2}{c}{JPEG} & \multicolumn{2}{c}{\textup{JP2K}}& \multicolumn{2}{c}{\textup{GB}}& \multicolumn{2}{c}{\textup{GN}} &\multicolumn{2}{c}{\textup{Overall}}\\ \hline
         & PLCC & SRCC  & PLCC & SRCC  & PLCC & SRCC  & PLCC & SRCC  & PLCC & SRCC  & PLCC & SRCC & PLCC & SRCC  & PLCC & SRCC \\ \hline

         S-PSNR & 0.151 & -0.113 & 0.931 & 0.890 & 0.225 & -0.103 &
         0.890 & 0.847 & 0.886 & 0.887 & 0.784 & 0.780 & 0.915 & 0.881 & 0.763 & 0.751\\

         S-SSIM & 0.149 & 0.055 & 0.922 & \textbf{0.932} & 0.018 & -0.031
         & 0.922 & 0.903 & 0.930 & 0.931 & 0.869 & 0.870 & 0.955 & 0.941 & 0.828 & 0.823 \\

         WS-PNSR & 0.153 & -0.116 & 0.931 & 0.893 & 0.215 & -0.104
         & 0.890 & 0.847 & 0.886 & 0.886 & 0.785 & 0.781 & 0.915 & 0.881 & 0.764 & 0.751\\

         CPP-PNSR & 0.129 & -0.054 & 0.930 & 0.906 & 0.215 & -0.079
         & 0.891 & 0.849 & 0.885 & 0.885 & 0.767 & 0.764 & 0.914 & 0.878 & 0.757 & 0.747 \\  \hline

         PSNR &0.165 & -0.114 & 0.924 & 0.893 & 0.231 & -0.102
         & 0.891 & 0.848 & 0.891 & 0.893 & 0.759 & 0.754 & 0.925 & 0.895 & 0.744 & 0.733 \\

         V-PSNR & 0.148 & -0.049 & 0.928 & 0.893 & 0.241 & -0.077
         & 0.905 & 0.898 & 0.897 & 0.896 & 0.835 & 0.831 & 0.913 & 0.884 & 0.795 & 0.779 \\

         O-PSNR & 0.583 & 0.516 & \textbf{0.933} & 0.911 & 0.597 & 0.467
         & 0.905 & 0.891 & 0.901 & 0.901 & 0.884 & 0.886 & 0.914 & 0.881 & 0.797 & 0.780\\  \hline

         SSIM & 0.148 & 0.057 & 0.910 & \textbf{0.932} & 0.036 & -0.030
         & 0.910 & 0.893 & 0.924 & 0.926 & 0.849  & 0.845 & 0.951 & 0.937 & 0.809 & 0.802\\

         V-SSIM & 0.149 & 0.044 & 0.930 & 0.916 & 0.038 & -0.038
         & 0.924 & 0.905 & 0.932 & 0.931 & 0.891 & 0.891 & 0.942 & 0.929 & 0.850 & 0.844 \\

         O-SSIM & 0.468 & 0.495 & 0.923 & 0.881 & 0.579 & 0.435
         & 0.938 & 0.922 & 0.941 & 0.939 & 0.918 & 0.921 & 0.942 & 0.930 & 0.866 & 0.862 \\\hline

         VIF & 0.111 & 0.057 & 0.920 & 0.872 & 0.356 & 0.331
         & 0.916 & 0.900 & 0.955 & 0.956 & 0.960 & 0.958 & 0.950 & 0.921 & 0.871 & 0.862\\

         V-VIF & 0.151 & 0.046 & 0.923 & 0.861 & 0.493 & 0.342
         & 0.929 & 0.915 & 0.960 & 0.962 & 0.957 & 0.954 & 0.947 & 0.916 & 0.883 & 0.873 \\

         O-VIF & \textbf{0.605} & \textbf{0.555} & 0.893 & 0.843 & 0.617 & 0.496
         & 0.937 & 0.923 & 0.969 & 0.968 & 0.965 & 0.965 & 0.947 & 0.917 & 0.889 & 0.880 \\ \hline

         NLPD & 0.012 & -0.009 & 0.907 & 0.870 & 0.244 & -0.063
         & 0.925 & 0.945 & 0.919 & 0.947 & 0.849  & 0.893 & 0.952 & 0.947 & 0.854 & 0.844 \\

         V-NLPD & 0.069 & -0.017 & 0.895 & 0.892 & 0.244 & -0.065
         & 0.964 & 0.954 & 0.954 & 0.954 & 0.933 & 0.933 & 0.970 & 0.957 & 0.911 & \textbf{0.907}\\

         O-NLPD & 0.479 & 0.534 & 0.898 & 0.857 & 0.311 & 0.472
         & \textbf{0.972} & \textbf{0.958} & 0.964 & 0.962 & 0.942 & 0.945 & \textbf{0.974} & \textbf{0.963} & \textbf{0.912} & \textbf{0.907} \\\hline

         DISTS & 0.079 & 0.025 & 0.867 & 0.861 & 0.450 & 0.299
         & 0.863 & 0.915 & 0.939 & 0.952 & 0.959 & 0.956 & 0.951 & 0.944 & 0.837 & 0.830 \\

         V-DISTS & 0.055 & 0.069 & 0.900 & 0.910 & 0.512 & 0.402
         & 0.942 & 0.937 & 0.961 & 0.959 & 0.965 & 0.957 & 0.963 & 0.949 & 0.883 & 0.875\\

         O-DISTS & 0.489 & 0.518 & 0.916 & 0.903 & \textbf{0.660} & \textbf{0.613}
         & 0.955 & 0.942 & \textbf{0.971} & \textbf{0.969} & \textbf{0.973} & \textbf{0.969} & 0.966 & 0.952 & 0.882 & 0.875 \\

         \bottomrule
    \end{tabular}
    \caption{Performance comparison of omnidirectional IQA methods on the proposed and OIQA databases. The best results are highlighted in bold.}
    \label{tab:performance_proposedDatabase_OIQA}
\end{table*}

\begin{table*}[!htb]
  \renewcommand\arraystretch{1.05}
  \small%
	\centering%
  \begin{tabular}{  r| c c c c c c c c c c c c c c }
  \toprule
  & \multicolumn{2}{c}{GB} & \multicolumn{2}{c}{GN} & \multicolumn{2}{c}{ST} & \multicolumn{2}{c}{VP9} & \multicolumn{2}{c}{H.265} & \multicolumn{2}{c}{DS} & \multicolumn{2}{c}{Overall} \\ \hline
  & \textup{PLCC} & \textup{SRCC}  & \textup{PLCC} & \textup{SRCC}  & \textup{PLCC} & \textup{SRCC}  & \textup{PLCC} & \textup{SRCC}  & \textup{PLCC} & \textup{SRCC}  & \textup{PLCC} & \textup{SRCC}  & \textup{PLCC} & \textup{SRCC} \\ \hline
  S-PSNR & 0.887 & 0.758 & 0.897 & 0.863 & 0.657 & 0.638 & 0.608 & 0.594 & 0.806 & 0.797 & 0.865 & 0.780 & 0.727 & 0.649 \\
  S-SSIM & 0.902 & 0.810 & 0.920 & 0.905 & 0.634 & 0.612 & 0.784 & 0.756 & 0.874 & 0.874 & 0.853 & 0.778 & 0.746 & 0.722 \\
  WS-PSNR & 0.887 & 0.758 & 0.897 & 0.863 & 0.656 & 0.634 & 0.607 & 0.595 & 0.806 & 0.797 & 0.865 & 0.779 & 0.727 & 0.649 \\
  CPP-PSNR &  0.885 & 0.750 & 0.898 & 0.879 & 0.635 & 0.615 & 0.596 & 0.582 & 0.800 & 0.787 & 0.866 & 0.776 & 0.725 & 0.646 \\  \hline
  PSNR & 0.877 & 0.730 & 0.896 & 0.857 & 0.623 & 0.618 & 0.572 & 0.561 & 0.780 & 0.766 & 0.863 & 0.786 & 0.715 & 0.642  \\
  V-PSNR & 0.901 & 0.792 & 0.897 & 0.861 & 0.689 & 0.684 & 0.662 & 0.649 & 0.836 & 0.829 & 0.877 & 0.802 & 0.753 & 0.689 \\
  O-PSNR & 0.916 & 0.841 & 0.893 & 0.848 & 0.704 & 0.693 & 0.744 & 0.722 & 0.884 & 0.888 & 0.882 & 0.812 & 0.758 & 0.672
\\  \hline
  SSIM & 0.891 & 0.781 & 0.913 & 0.898 & 0.551 & 0.542 & 0.770 & 0.743 & 0.848 & 0.844 & 0.860 & 0.759 & 0.729 & 0.704 \\
  V-SSIM & 0.911 & 0.830  & 0.925 & 0.911 & 0.643 & 0.622 & 0.809 & 0.785 & 0.899 & 0.903 & 0.870 & 0.804 & 0.768 & 0.749 \\
  O-SSIM & 0.920 & 0.863 & 0.924 & 0.911 & 0.652 & 0.640 & 0.822 & 0.785 & 0.927 & 0.930 & 0.892 & 0.825 & 0.769 & 0.716
\\ \hline
  VIF & 0.943 & 0.893 & 0.927 & 0.922 & 0.691 & 0.694 & 0.839 & 0.819 & 0.913 & 0.918 & 0.904 & 0.806 & 0.841 & 0.833 \\
  V-VIF & 0.944 & 0.899 & 0.924 & 0.919 & \textbf{0.754} & 0.754 & 0.862 & 0.841 & 0.926 & 0.930 & 0.904 & 0.812 & 0.845 & 0.839 \\
  O-VIF & 0.951 & 0.914 & 0.917 & 0.905 & 0.740 & 0.754 & 0.877 & \textbf{0.873} & 0.941 & 0.946 & 0.904 & 0.814 & 0.790 & 0.753 \\  \hline
  NLPD & 0.923 & 0.860 & 0.923 & 0.913 & 0.659 & 0.646 & \textbf{0.878} & 0.861 & 0.930 & 0.931 & 0.881 & 0.790 & 0.653 & 0.621 \\
  V-NLPD & 0.937 & 0.887 & \textbf{0.935} & \textbf{0.930} & 0.707 & 0.701 & 0.872 & 0.849 & 0.945 & 0.951 & 0.886 & 0.805 & 0.705 & 0.671 \\
  O-NLPD & 0.940 & 0.895 & 0.931 & 0.920 & 0.721 & 0.719 & 0.866 & 0.845 & \textbf{0.946} & \textbf{0.952} & 0.894 & 0.813 & 0.735 & 0.701  \\  \hline
  DISTS & 0.954 & 0.921 & 0.921 & 0.910 & 0.664 & 0.649 & 0.738 & 0.708 & 0.854 & 0.855 & 0.920 & 0.852 & 0.749 & 0.748 \\
  V-DISTS & 0.954 & 0.917 & 0.903 & 0.890 & 0.764 & \textbf{0.758} & 0.765 & 0.749 & 0.869 & 0.872 & 0.947 & 0.920 & 0.814 & 0.813 \\
  O-DISTS & \textbf{0.958} & \textbf{0.926} & 0.900 & 0.888 & 0.708 & 0.703 & 0.802 & 0.781 & 0.895 & 0.890 & \textbf{0.950} & \textbf{0.938} & \textbf{0.850} & \textbf{0.851} \\
  \bottomrule
  \end{tabular}%
  \caption{PLCC and SRCC results of OIQA methods on the LIVE database.}
  \label{tab:performance_3D_PLCC_SRCC}
\end{table*}

\section{Experiments}
In this section, we first describe the implementation details of the proposed computational framework for omnidirectional IQA. Next, we introduce the evaluation procedures, and compare our methods with state-of-the-art quality measures, followed by a statistical significance test. Last, we conduct comprehensive ablation studies to analyze the sensitivity of individual components.

\subsection{Implementation Details and Evaluation Protocols}
\begin{figure}[htb]
 \centering
 \includegraphics[width=\columnwidth]{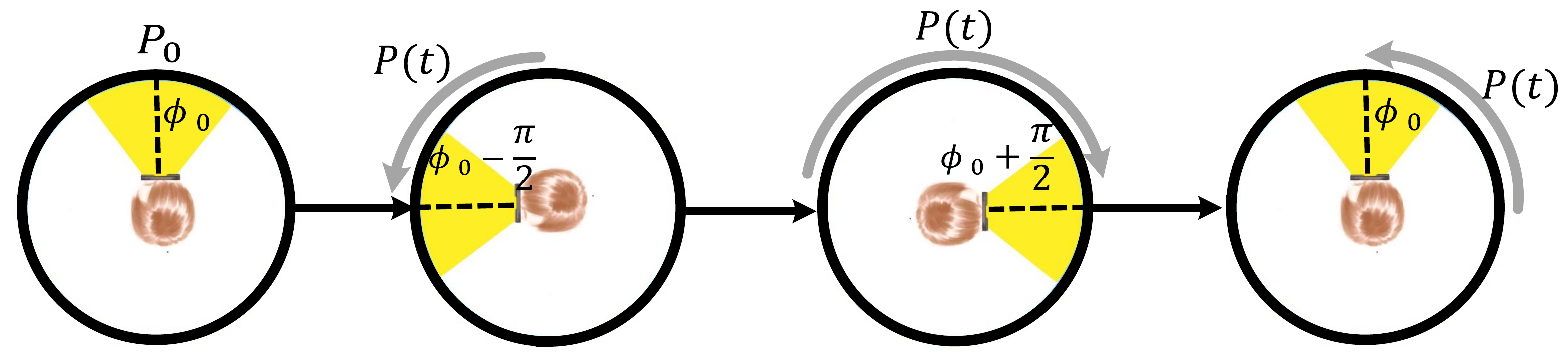}
 \caption{Illustration of the default scanpath. $\phi_0$ is the longitude of the starting point.}
 \label{fig:scanpath_example}
\end{figure}
The proposed computational framework requires user viewing conditions and behaviors to transform static panoramas to moving camera videos.
When such information is not available, the overall quality score may be obtained by taking the empirical expectation over several preferable types of viewing conditions and behaviors. Specifically, we sample four different starting points evenly spaced along the equator.
Considering a reasonable gaze speed of $24\degree/\mathrm{s}$~\cite{Sitzmann2018Saliency_Analyze}, we set a fixed exploration time $T$ to $15$ seconds \cite{Upenik2016_subjective,Huang2018_subjective,Sun2017_subjective,Chen2020_subjective,Duan2018_subjective}. To keep the computational complexity manageable, we design a single scanpath by taking into account the fact that the front equator regions are viewed more frequently than other parts. Specifically, the user first browses the panorama from the starting point $(\phi_0, 0)$, then gradually moves the gaze counterclockwise along the equator to $(\phi_0-\pi/2, 0)$ for viewing the left part of the 360{\degree} image. Next, the user begins to explore the right part of the scene by moving the gaze clockwise  from $(\phi_0-\pi/2, 0)$ to $(\phi_0+\pi/2, 0)$. Finally, the user returns to the starting point $(\phi_0, 0)$ and finishes the browsing (see \autoref{fig:scanpath_example}). Note that we constrain the gaze movements along the equator by clamping the latitude to  $\theta=0$. The detailed specifications of the default viewing conditions and behaviors are summarized in \autoref{tab:default_setting}.

Given a pair of reference and distorted panoramas, we first downsample them to reduce the computational complexity as suggested in \cite{Zhou2004SSIM_IQA}.
The implementations of the five full-reference IQA models are obtained from the respective authors. The three parameters in the temporal hysteresis model, including the memory duration $K=20$, the normalized Gaussian weighting function $w$ with standard deviation  $(2K-1)/12$, and the linear factor $\alpha=0.8$, are set according to \cite{Seshadrinathan2011TH_VQA}.

 We use three subject-rated VR datasets - the proposed database in Section \ref{sec:st}, the OIQA database in \cite{Duan2018_subjective}, and the LIVE 3D VR IQA database (LIVE) in \cite{Chen2020_subjective}. The OIQA database contains $320$ distorted panoramas, generated from $16$ reference panoramas with four distortion types at five distortion levels, including JPEG compression (JPEG), JPEG2000 compression (JP2K), Gaussian noise (GN), and Gaussian blur (GB). The LIVE database includes $15$ reference stereoscopic omnidirectional panoramas. Six distortion types with five levels are applied to produce $450$ distorted images, including GN, GB, downsampling (DS), stitching distortion (ST), VP9 compression, and H.265 compression.

 We use two evaluation metrics to quantify the quality prediction performance, including PLCC and Spearman’s rank-order correlation coefficient (SRCC).  A better quality model achieves higher PLCC and SRCC values. As suggested in \cite{VQGE_params}, we map model predictions to human quality ratings through a four-parameter logistic function before calculating PLCC:
\begin{align}
    f(Q) = (\beta_1-\beta_2)\frac{1}{1+e^{-\frac{Q-\beta_3}{\left|\beta_4\right|}}}+\beta_2,
\end{align}
where $\{\beta_i\}_{i=1}^4$ are the parameters to be fitted.

\begin{figure*}[!htb]
    \centering
    \subfloat[Proposed database]{\includegraphics[width=0.32\textwidth]{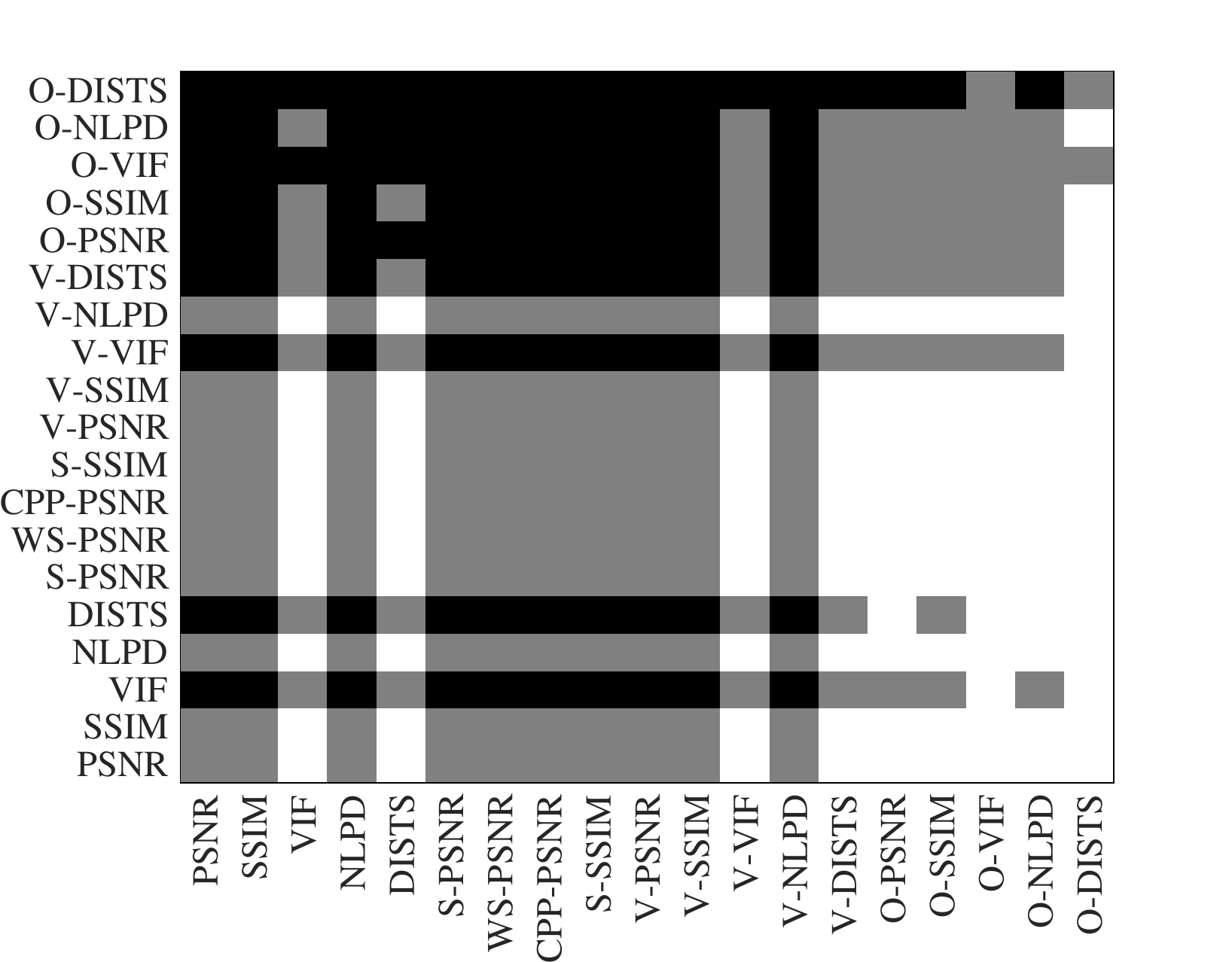}}
    \subfloat[OIQA]{\includegraphics[width=0.32\textwidth]{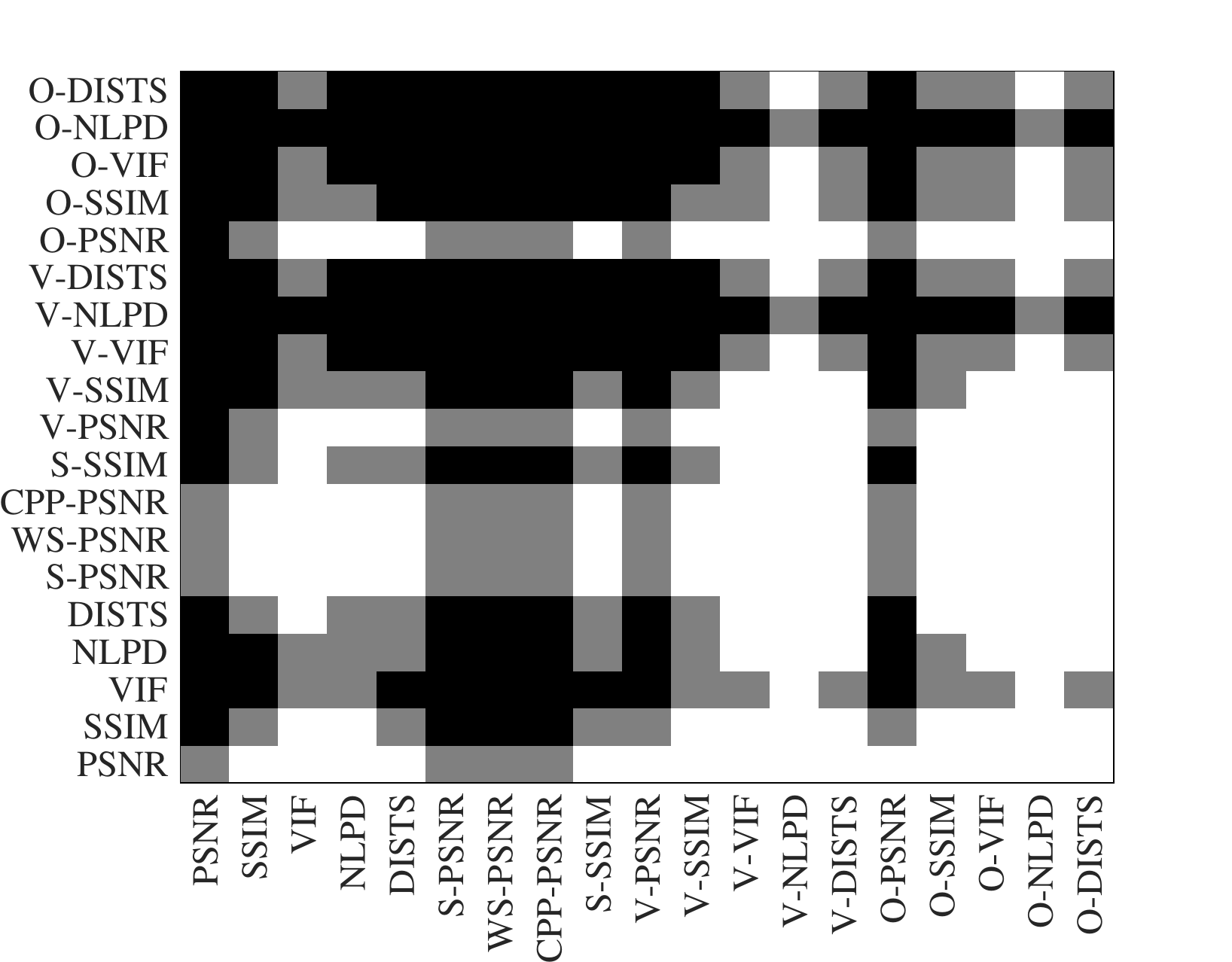}}
    \subfloat[LIVE]{\includegraphics[width=0.32\textwidth]{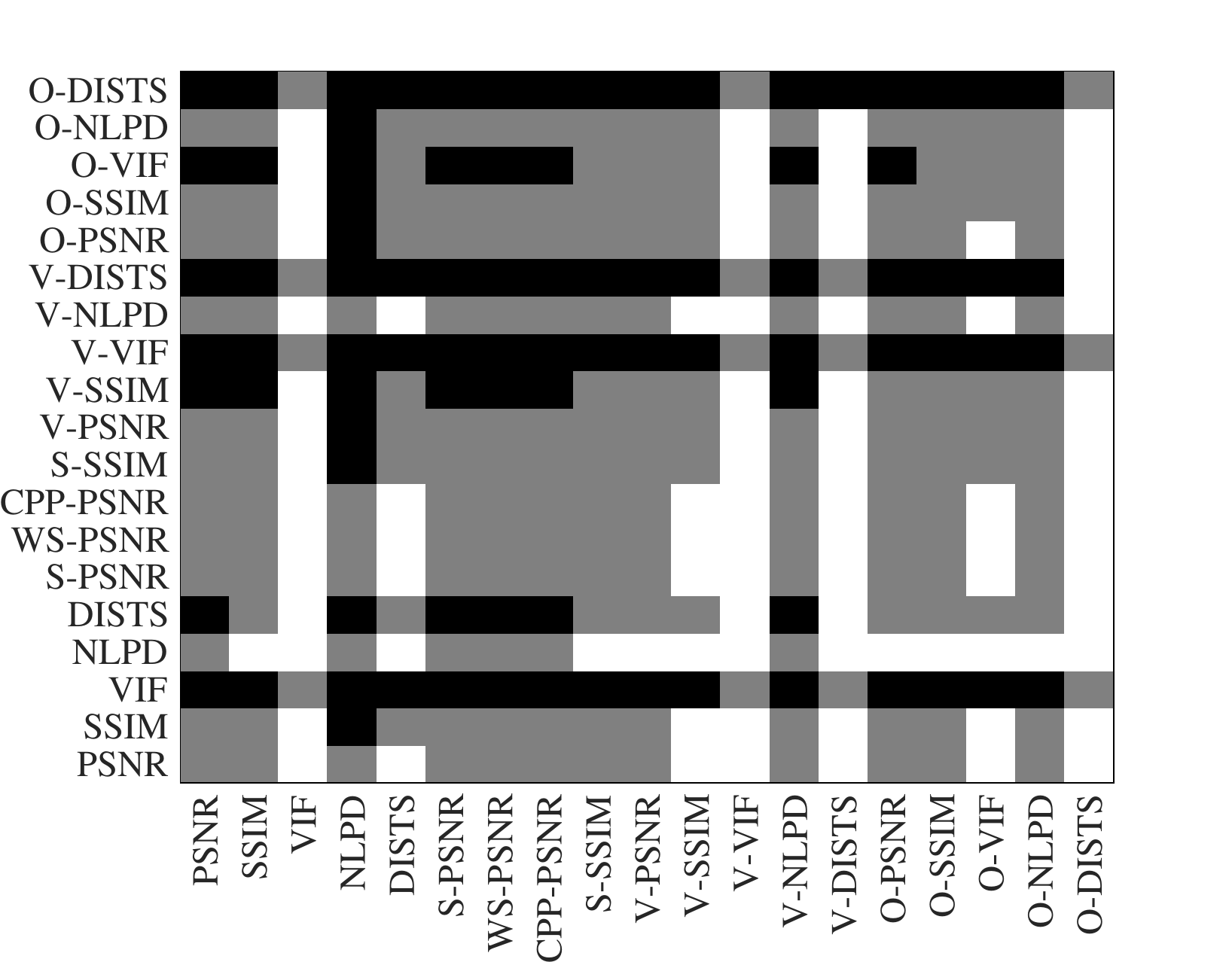}}
    \caption{Statistical significance diagram based on quality prediction residuals using $F$-test. A black block means the row model performs significantly better than the column model, a white block means the opposite, and a gray block indicates the significant difference between the row and column models is not observed.}
    \label{fig:significance_test}
\end{figure*}

\subsection{Main Results}
We add an ``O-'' to the five 2D IQA methods listed in Section \ref{subsec:ins} as a prefix to name the proposed quality models (\eg, PSNR to O-PSNR). We include S-PSNR\cite{Yu2015SPSNR_OIQA}, S-SSIM \cite{Lopes2018WSSIM_OIQA}
, WS-PSNR \cite{Sun2017WSPSNR_OIQA}, and CPP-PSNR \cite{Zakharchenko2016CPPPSNR_OIQA} as representative omnidirectional IQA models for comparison. We also directly apply 2D IQA models to equirectangular projections as baselines. To measure relative perceptual gains when incorporating viewing conditions and behaviors, we further create a set of viewport-based methods by extracting viewports uniformly distributed over the sphere for quality computation using the same 2D IQA models, as suggested in\cite{Xu2019_viewports}. Similarly, we add a ``V-'' in front to name viewport-based methods.
For all models, we compute quality values on panoramas of the same downsampled resolution. The quality score of a stereoscopic image is computed by averaging quality estimates of both views. We list the quantitative results on the proposed database and the OIQA database in \autoref{tab:performance_proposedDatabase_OIQA}, and the LIVE database in Table \ref{tab:performance_3D_PLCC_SRCC}, from which we make several interesting observations.

First, it is quite surprising that recent 2D IQA models directly applied to equirectangular projections outperform existing omnidirectional IQA models. For example, the performance of S-PSNR, S-SSIM, WS-PSNR, and CPP-PSNR is worse than that of VIF, NLPD, and DISTS on the OIQA database \cite{Duan2018_subjective}. This suggests that instead of sticking to standard quality measures - PSNR and SSIM, we may transfer recent advances in the domain of 2D IQA to VR applications. Second, current omnidirectional IQA methods, projection-based methods, and viewport-based methods fail to capture the \textit{localized} stitching distortions in the proposed database. For example, the SRCC values of PSNR, S-PSNR, and V-PSNR are even negative (see \autoref{tab:performance_proposedDatabase_OIQA}). When the stitching distortions are synthesized \textit{globally} as in the LIVE database, the overall results get back to a reasonable level (see Table \ref{tab:performance_3D_PLCC_SRCC}). Finally, the proposed computational framework achieves significant performance improvements compared to both projection-based and viewport-based methods, especially on the proposed database.  O-NLPD and O-DISTS also achieve the best performance on the OIQA database and the LIVE database, respectively. A noticeable exception is O-VIF, which underperforms VIF and V-VIF on the LIVE database. This may be because VIF tends to over-penalize the stitching distortions, which have large differences in pixel values but look more acceptable compared to other distortion types. The temporal hysteresis model in O-VIF tends to amplify such penalties, resulting in a significant performance drop.

\begin{table}[!b]
  \renewcommand\tabcolsep{6pt}
  \small%
	\centering%
  \begin{tabular}{ c | c | c c c c c c}
  \toprule
   & & AM & HM & GW & MM & PS & \textbf{TH} \\ \hline
   \multirow{2}*{O-PSNR} & PLCC & 0.565 & 0.628 & \textbf{0.767} & 0.538 & 0.586 & 0.597 \\
   & SRCC & 0.470 & 0.449 & \textbf{0.737} & 0.453 & 0.307 & 0.467 \\ \hline
   \multirow{2}*{O-SSIM} & PLCC & 0.573 & \textbf{0.594} & 0.470 & 0.577 & 0.524 & 0.579 \\
   & SRCC & 0.417 & 0.396 & 0.525 & 0.429 & \textbf{0.527} & 0.435 \\  \hline
   \multirow{2}*{O-DISTS} & PLCC & 0.681 & 0.677 & \textbf{0.843} & 0.694 & 0.682 & 0.660 \\
   & SRCC & 0.626 & 0.620 & \textbf{0.844} & 0.648 & 0.584 & 0.613 \\
  \bottomrule
  \end{tabular}%
   \caption{\centering Performance comparison of our models with different temporal pooling strategies on the proposed database. AM: arithmetic mean. HM: harmonic mean. GW: ascending half of Gaussian weighting. MM: Minkowski mean. PS: percentile scoring.  The default pooling is highlighted in bold.}
    \label{tab:comparsion_ablation}
\end{table}

\begin{table}[!hb]
  \small%
	\centering%
  \begin{tabular}{ c | c | c| c c  }
  \toprule
   & Input size & Viewport size& PLCC& SRCC\\ \hline
   \multirow{3}*{O-PSNR} & 1,920  $\times$ 3,840 & 640 $\times$ 640 & 0.573 & 0.454  \\
  & \textbf{960 $\times$ 1,920} & \textbf{320 $\times$ 320} & 0.597 & 0.467  \\
  & 480 $\times$ 960 & 160 $\times$ 160 & \textbf{0.603} &\textbf{0.470}  \\ \hline

  \multirow{3}*{O-SSIM} & 1,920  $\times$ 3,840 & 640 $\times$ 640 & \textbf{0.601} & 0.399
  \\ & \textbf{960 $\times$ 1,920} & \textbf{320 $\times$ 320} & 0.579 & \textbf{0.435}  \\
  & 480 $\times$ 960 & 160 $\times$ 160 & 0.506 & 0.351 \\ \hline

  \multirow{3}*{O-DISTS} & 1,920  $\times$ 3,840 & 640 $\times$ 640 & 0.647 & 0.602
  \\ & \textbf{960 $\times$ 1,920} & \textbf{320 $\times$ 320} & 0.660 & 0.613 \\
  & 480 $\times$ 960 & 160 $\times$ 160 & \textbf{0.667} & \textbf{0.624}  \\
  \bottomrule
  \end{tabular}%
  \caption{Performance comparison of our models with different input resolutions on the proposed database. The default size determined by \cite{Zhou2004SSIM_IQA} is highlighted in bold.}
  \label{tab:comparsion_differentSize}
\end{table}

To ascertain that the improvement of the proposed computational framework is statistically significant, we carry out a statistical significance test by following the approach introduced in \cite{Sheikh2006}. First, a nonlinear function is applied to map objective quality scores to subjective scores. We observe that the prediction residuals all have zero mean, and thus the model with a lower variance is generally considered better. We conduct a hypothesis testing using the $F$-statistic, \ie,  the ratio of variances. The null hypothesis is that the prediction residuals of one quality model come from the same distribution, and are statistically indistinguishable (with $95\%$ confidence) from the residuals of another model. After comparing every possible pairs of objective models, the results are summarized in \autoref{fig:significance_test}, where a black block means the row model performs significantly better than the column model, a white block means the opposite, and a gray block indicates the significant difference between the row and column models is not observed. From the figure, we conclude that quality models within the proposed computational framework are statistically better than the competing methods in most cases.

\subsection{Ablation Experiments}
In this subsection, we conduct a series of ablation experiments to analyze the impact of temporal pooling strategies, input resolutions, sampling rates, and scanpath patterns within the proposed computational framework. Here we only consider three 2D IQA measures as base models: PSNR, SSIM, and DISTS.

\paragraph{Choice of Temporal Pooling Strategy} In addition to the temporal hysteresis pooling \cite{Seshadrinathan2011TH_VQA}, we test another five strategies, including arithmetic mean, harmonic mean, the ascending half of Gaussian weighting \cite{Seshadrinathan2011TH_VQA}, Minkowski mean (using $\ell_2$-norm) and percentile scoring (using $10\%$) \cite{Moorthy2009_Percentile}. From \autoref{tab:comparsion_ablation}, we
find that temporal pooling makes a noticeable difference on the proposed database. For example, when DISTS \cite{Ding2020DISTS_IQA} is the base model,  switching the default hysteresis pooling to Gaussian weighting significantly boosts the performance, better accounting for the recency effect. This verifies our omnidirectional image-to-video conversion as a natural way of incorporating viewing conditions and behaviors into the quality assessment process.

\begin{table}[!t]
  \small%
	\centering%
  \begin{tabular}{ c | c | c| c c c }
  \toprule
  & $s_1$ & $R$ & PLCC & SRCC \\ \hline
   \multirow{3}*{O-PSNR}
   & 4 & 5 & \textbf{0.599} & 0.466  \\
  & 2 & 10 & 0.597 & 0.466 \\
  & 1 & \textbf{20} & 0.597 & \textbf{0.467} \\ \hline

  \multirow{3}*{O-SSIM}
  & 4 & 5 & 0.574 & 0.379  \\
  & 2 & 10 & 0.574 & 0.379\\
  & 1 & \textbf{20} &\textbf{0.579} & \textbf{0.435} \\\hline

  \multirow{3}*{O-DISTS}
  & 4 & 5 & \textbf{0.661} & \textbf{0.613}  \\
  & 2 & 10 & 0.660 & \textbf{0.613}  \\
  & 1 & \textbf{20} & 0.660 & \textbf{0.613}  \\
  \bottomrule
  \end{tabular}%
  \caption{Performance comparison of our models with different sampling rates on the proposed database. The sampling rate supported by the HMD is highlighted in bold.}
  \label{tab:comparsion_sampleRate}
\end{table}

\paragraph{Choice of Input Resolution} The resolution of the input 360{\degree} images determines the effective viewing distance and the sizes of viewport.  \autoref{tab:comparsion_differentSize} shows the results on the proposed database, where we observe that model performance is generally better as the input resolution reduces.
In our implementation, we employ automatic downsampling as suggested in \cite{Zhou2004SSIM_IQA} to keep the shorter side of the panorama in the range of $512$ and $1024$. Accordingly, the size of the square viewport is in the range of $170\times 170$ and $341\times 341$. As such, we strike a good balance between signal fidelity and computational complexity.

\begin{table}[!t]
  \small%
	\centering%
  \begin{tabular}{ c | l| c c c }
  \toprule
  & Scanpath pattern & PLCC & SRCC \\ \hline
   \multirow{3}*{O-PSNR}
   & Default scanpath & \textbf{0.797} & \textbf{0.780}  \\
    & Default scanpath with nonzero latitudes & \textbf{0.797} & \textbf{0.780} \\
  & Counterclockwise rotation & \textbf{0.797} & \textbf{0.780} \\ \hline

  \multirow{3}*{O-SSIM}
    & Default scanpath & \textbf{0.866} & \textbf{0.862}  \\
      & Default scanpath with nonzero latitudes & \textbf{0.866} & 0.861 \\
  & Counterclockwise rotation & 0.865 & 0.861 \\ \hline

  \multirow{3}*{O-DISTS}
  & Default scanpath & 0.882 &  \textbf{0.875}  \\
  & Default scanpath with nonzero latitudes & 0.882 & \textbf{0.875}  \\
  & Counterclockwise rotation & \textbf{0.883} &  \textbf{0.875} \\
  \bottomrule
  \end{tabular}%
  \caption{Performance comparison of our models with different scanpath patterns on the OIQA database, where users' scanpaths are not available.}
  \label{tab:comparsion_movement}
\end{table}

\paragraph{Choice of Sampling Rate}
In the proposed computational framework, we generate video sequences by sampling viewports along users' scanpaths at a certain rate. Thus, it is natural to ask: what is the optimal sampling rate in terms of prediction accuracy and computational complexity?
We test our models with different sampling rates by adjusting the stride parameter $s_1$ in Eq. \eqref{eq:sr}, and list the results on the proposed database in \autoref{tab:comparsion_sampleRate}. We find that they are robust to variation of sampling rates. In our default setting, a constant sampling rate is assumed to extract viewports that are uniformly distributed along the scanpath. However, in practice, humans tend to alternate between two modes: attention and re-orientation \cite{Sitzmann2018Saliency_Analyze}. The attention mode is activated when viewers have paused on interesting parts of the scene, while the re-orientation mode begins when the human eye moves to new salient regions. In the future, we may take advantage of this viewing behavior for computational complexity reduction.

\paragraph{Choice of Scanpath Pattern} To investigate the impact of scanpaths on the prediction accuracy, we test two additional trajectories within the proposed computational framework: 1) default scanpath with nonzero latitudes by adding a random Brownian motion to the current latitude and 2) counterclockwise rotation along the equator for $2\pi$. The results on the OIQA database are listed in~\autoref{tab:comparsion_movement}, from which we can see that no scanpath pattern seems to be significantly better than the others. The reason may be that most test panoramas are distorted globally in the OIQA database, resulting in relatively uniform quality.

\section{Conclusion and Discussion}
In this paper, we have taken steps towards perceptual quality assessment of omnidirectional images. We conducted a psychophysical experiment to study how the viewing conditions (\ie, the starting point and the exploration time) affect user viewing behaviors and the perceived quality of 360{\degree} images.  We then introduced a computational framework to design objective omnidirectional IQA models, which incorporates the viewing conditions and behaviors into the quality prediction process. The key idea is to map panoramas to moving camera videos by extracting the sequences of viewports along the scanpaths. Experimental results on three VR IQA databases demonstrated the promise of the proposed framework, where we successfully transferred the advances in 2D IQA to VR applications.

Our framework suggests a natural extension to \textit{personalized} omnidirectional IQA, which may be more suitable in VR applications as user viewing behaviors tend to vary based on their own personal experiences and preferences. This can be easily achieved by exploiting behavior statistics of a single user, instead of averaging across several users.

Although the proposed methods have offered significant perceptual gains compared to existing models, they are somewhat limited at handling localized stitching distortions (as shown in \autoref{tab:performance_proposedDatabase_OIQA}). Although our omnidirectional image to video conversion may detect the stitching distortions as long as they are seen by viewers, this does not necessarily mean that existing objective quality models can quantify them in a proper way due to the idiosyncratic visual appearances (see \autoref{fig:example_stitching}).
We believe that current and future models that take better account for local distortions will have great potential in boosting the performance within the proposed computational framework.

The current work focuses on omnidirectional image quality with a fixed display constraint. However, what and how to display a panoramic image may have an impact on its perceived quality. For example, Zhang~\etal~\cite{Yingxue2018} suggested to resample the omnidirectional image to the optimal resolution before coding, as a way of matching the resolutions between the HMD and the image. Jabar \etal~\cite{Jabar2020} found the perceived quality of 360{\degree} images has a dependency on the FoV of the extracted viewport, and the optimal FOV for viewing panoramas is 110{\degree}. All of these suggest to incorporate the display information as an additional viewing condition into the proposed computational framework. Moreover, it would be interesting to build computational models to predict the overall quality-of-experience of users when exploring 360{\degree} images, including VR discomfort and sickness \cite{Kim2018_Sickness}.

\acknowledgments{This work was supported in part by the National Key R\&D Program of China under Grant 2018AAA0100601, the National Natural Science Foundation of China under Grants 62071407 and 61822109, the Fok Ying Tung Education Foundation under Grant 161061,  the Jiangxi Natural Science Foundation of China under Grant 20202ACB202007, and the CityU APRC Grant (9610487).}

\bibliographystyle{IEEEtran}

\bibliography{TVCG_v2_publish}

\begin{wrapfigure}[9]{l}{0.25\linewidth}
\vspace{-14pt}
\includegraphics[width=1in,height=1.25in,clip,keepaspectratio]{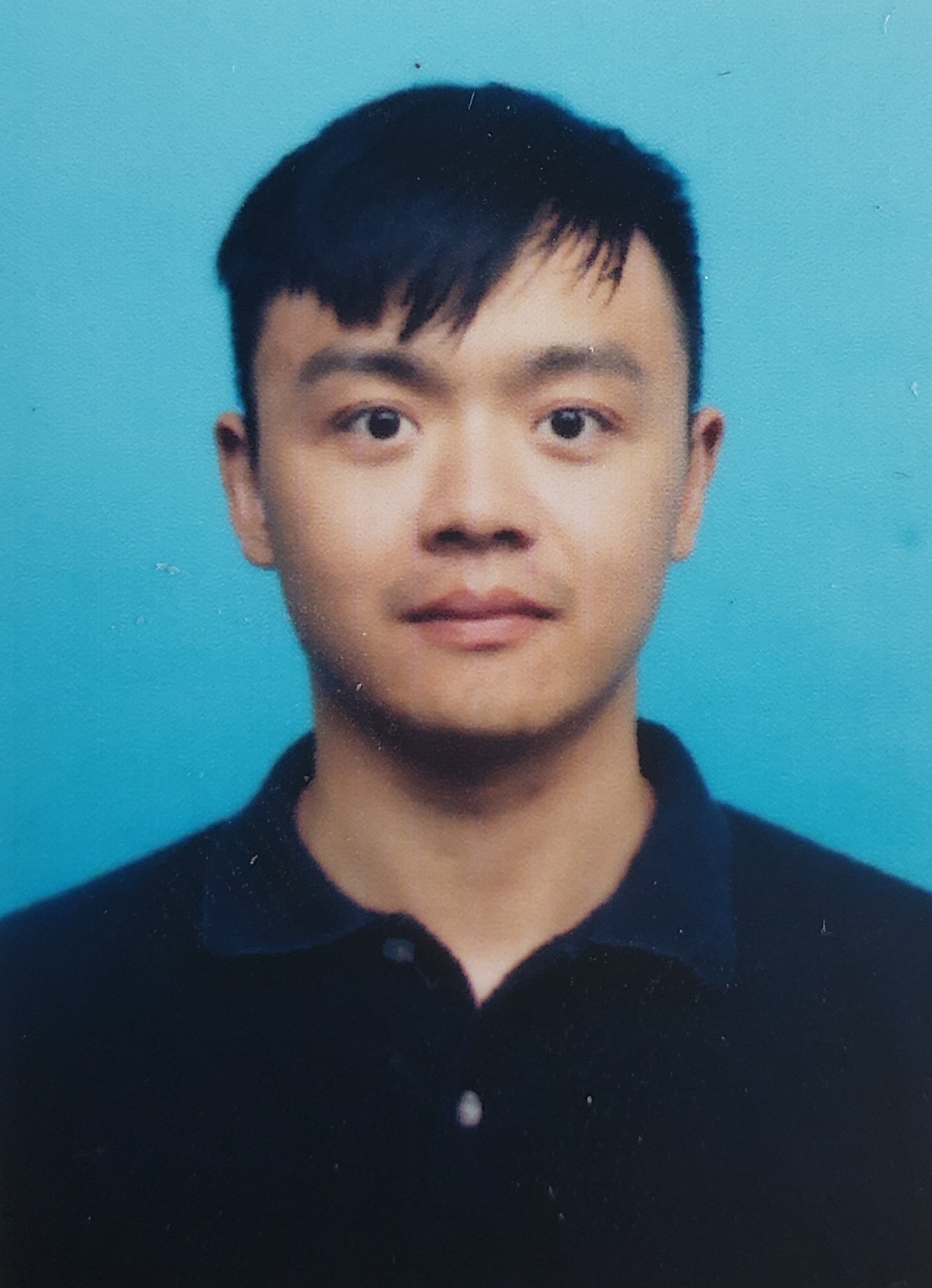}
\end{wrapfigure}
\noindent\textbf{\footnotesize {Xiangjie Sui}}
\footnotesize {received the B.E. degree from the Jiangxi University of Finance and Economics, Nanchang, China, in 2018. He is currently pursuing the M.A.Sc. degree with the School of Information Management, Jiangxi University of Finance and Economics, Nanchang, China. His research interests include visual quality assessment, and VR image/video processing.} \\ \qquad \\ \qquad \\

\begin{wrapfigure}[9]{l}{0.25\linewidth}
\vspace{-14pt}
\includegraphics[width=0.9in,height=1.25in,clip,keepaspectratio]{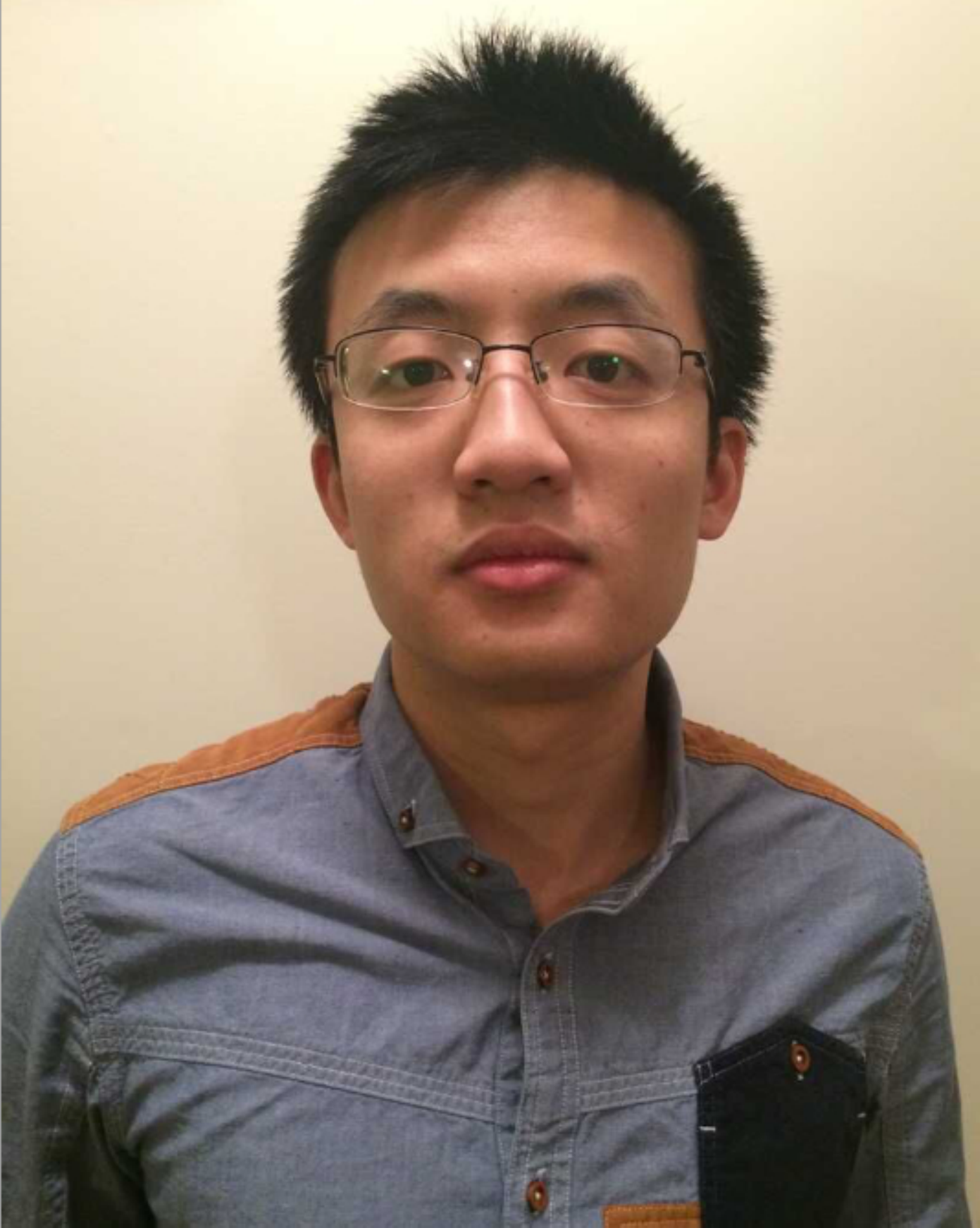}
\end{wrapfigure}
\noindent\textbf{\footnotesize {Kede Ma}}
\footnotesize {(S’13–M’18) received the B.E. degree from the University of Science and Technology of China, Hefei, China, in 2012, and the M.S. and Ph.D. degrees in electrical and computer engineering from the University of Waterloo, Waterloo, ON, Canada, in 2014 and 2017, respectively. He was a Research Associate with the Howard Hughes Medical Institute and New York University, New York, NY, USA, in 2018. He is currently an Assistant Professor with the Department of Computer Science, City University of Hong Kong. His research interests include perceptual image processing, computational vision, and computational photography.}\\

\begin{wrapfigure}[8]{l}{0.25\linewidth}
\vspace{-16pt}
\includegraphics[width=1in,height=1.25in,clip,keepaspectratio]{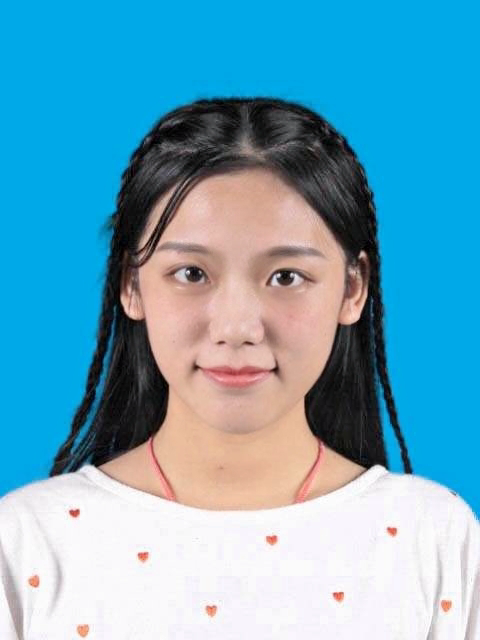}
\end{wrapfigure}
\noindent\textbf{\footnotesize {Yiru Yao}}
\footnotesize {received the B.E. degree from the Jiangxi University of Finance and Economics, Nanchang, China, in 2020. She is currently pursuing the M.A.Sc. degree with the School of Information Management, Jiangxi University of Finance and Economics, Nanchang, China. Her research interests include visual quality assessment, and VR image/video processing.}
\\ \qquad \\ \qquad \\

\begin{wrapfigure}[10]{l}{0.25\linewidth}
\vspace{-14pt}
\includegraphics[width=1in,height=1.25in,clip,keepaspectratio]{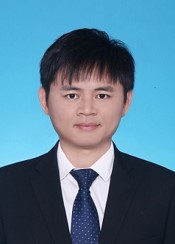}
\end{wrapfigure}
\noindent\textbf{\footnotesize {Yuming Fang}}
\footnotesize {(S’13–SM’17) received the B.E. degree from Sichuan University, Chengdu, China, the M.S. degree from the Beijing University of Technology, Beijing, China, and the Ph.D. degree from Nanyang Technological University, Singapore. He is currently a Professor with the School of Information Management, Jiangxi University of Finance and Economics, Nanchang, China. His research interests include visual attention modeling, visual quality assessment, computer vision, and 3D image/video processing. He serves as an Associate Editor for \textsc{IEEE ACCESS}. He serves on the Editorial Board of \textit{Signal Processing: Image Communication}.
}

\end{document}